\documentclass[fleqn,usenatbib]{mnras}

\usepackage[T1]{fontenc}
\usepackage{ae,aecompl}
\usepackage{graphicx}
\usepackage{amsmath}
\usepackage{amssymb}
\usepackage{longtable}
\usepackage{afterpage}
\usepackage{pdflscape}

\DeclareRobustCommand{\ion}[2]{%
\relax\ifmmode
\ifx\testbx\f@series
{\mathbf{#1\,\mathsc{#2}}}\else
{\mathrm{#1\,\mathsc{#2}}}\fi
\else\textup{#1\,{\mdseries\textsc{#2}}}%
\fi}

\newcommand{\Msun}{\hbox{$M_{\rm \odot}$}}

\newcommand\nodata{ ~$\cdots$~ }

\bibpunct{(}{)}{;}{a}{}{,}

\title[Intermediate redshift U\slash LIRGs]{Optical integral field spectroscopy of intermediate redshift infrared bright galaxies}

\author[M. Pereira-Santaella et al.]{\parbox{\textwidth}{M.~Pereira-Santaella$^{1}$\thanks{E-mail: miguel.pereira@physics.ox.ac.uk}, D.~Rigopoulou$^{1}$, G.~E.~Magdis$^{2}$,
N.~Thatte$^{1}$,
A.~Alonso-Herrero$^{3}$,
F.~Clarke$^{1}$,
D.~Farrah$^{4,5}$,
S.~Garc\'ia-Burillo$^{6}$,
L.~Hogan$^{1}$,
S.~Morris$^{7}$,
M.~Rodrigues$^{1}$,
J.-S.~Huang$^{8, 9, 10}$,
M.~Tecza$^{1}$
}
\vspace{0.4cm} \\
$^{1}$Department of Physics, University of Oxford, Keble Road, Oxford OX1 3RH, UK\\
$^{2}$Cosmic Dawn Center (DAWN), Niels Bohr Institute, University of  Copenhagen, Juliane Maries Vej 30, DK-2100 Copenhagen, Denmark\\
$^{3}$Centro de Astrobiolog\'ia (CAB, CSIC-INTA), ESAC Campus, E-28692 Villanueva de la Ca\~nada, Madrid, Spain\\
$^{4}$Department of Physics and Astronomy, University of Hawaii, 2505 Correa Road, Honolulu, HI 96822, USA\\
$^{5}$Institute for Astronomy, 2680 Woodlawn Drive, University of Hawaii, Honolulu, HI 96822, USA\\
$^{6}$Observatorio Astron\'omico Nacional (OAN-IGN)-Observatorio de Madrid, Alfonso XII, 3, E-28014, Madrid, Spain\\
$^{7}$Department of Physics \& Astronomy, University of British Columbia, Agricultural Road, Vancouver, BC, V6T 1Z1, Canada \\
$^{8}$National Astronomical Observatories of China, Chinese Academy of Sciences, Beijing 100012, China\\
$^{9}$China-Chile Joint Center for Astronomy, Chinese Academy of Sciences, Camino El Observatorio, 1515, Las Condes, Santiago, Chile\\
$^{10}$Harvard-Smithsonian Center for Astrophysics, 60 Garden Street, Cambridge, MA 02138, USA\\
}

\pubyear{2019}

\begin{document}
\label{firstpage}
\pagerange{\pageref{firstpage}--\pageref{lastpage}}
\maketitle

\begin{abstract}
The extreme infrared (IR) luminosity of local luminous and ultra-luminous IR galaxies (U\slash LIRGs; $11 <\log L_{\rm IR}\slash L_\odot < 12$ and $\log L_{\rm IR}\slash L_\odot>12$, respectively) is mainly powered by star-formation processes triggered by mergers or interactions. While U\slash LIRGs are rare locally, at $z>1$, they become more common, they dominate the star-formation rate (SFR) density, and a fraction of them are found to be normal disk galaxies. Therefore, there must be an evolution of the mechanism triggering these intense starbursts with redshift. To investigate this evolution, we present new optical SWIFT integral field spectroscopic H$\alpha$+[\ion{N}{ii}] observations of a sample of 9 intermediate-$z$ ($0.2<z<0.4$) U\slash LIRG systems selected from {\it Herschel} 250\micron\ observations. The main results are the following: 
(a) the ratios between the velocity dispersion and the rotation curve amplitude indicate that 10--25\% (1--2 out of 8) might be compatible with being isolated disks while the remaining objects are interacting\slash merging systems;
(b) the ratio between un-obscured and obscured SFR traced by H$\alpha$ and $L_{\rm IR}$, respectively, is similar in both local and these intermediate-$z$ U\slash LIRGs;
{and (c) the ratio between 250\micron\ and the total IR luminosities of these intermediate-z
U\slash LIRGs is higher than that of local U\slash LIRGs with the same $L_{\rm IR}$. This indicates a reduced dust temperature in these intermediate-$z$ U\slash LIRGs.} This, together with their already measured enhanced molecular gas content, suggests that the interstellar medium conditions are different in {our sample of} intermediate-$z$ galaxies when compared to local U\slash LIRGs.
\end{abstract}


\section{Introduction}\label{s:intro}

Discovered in large numbers by the Infrared Astronomical Satellite (\textit{IRAS}), luminous ($11<\log L_{\rm IR(8-1000\mu m)}\slash L_{\odot} <12$; LIRGs) and ultra-luminous infrared galaxies ($\log L_{\rm IR(8-1000\mu m)}\slash L_\odot>12$; ULIRGs) are amongst the most intensely star-forming galaxies in the Universe. While scarce locally, they are far more numerous at high redshifts and are responsible for about 50\% of the total star-formation rate (SFR) density at $z>1$ (e.g., \citealt{LeFloch2005, PerezGonzalez2005, Caputi2007, Magnelli2011}). As such, U\slash LIRGs represent an important population for understanding galaxy evolution.

In the local Universe, U\slash LIRGs are primarily found in interacting and$/$or merging systems. They exhibit a wide range of morphologies suggesting different dynamical phases ranging from isolated disks in low-luminosity LIRGs to strongly interacting galaxies and fully merged systems \citep{Kim1998, Kim2013, Farrah2001, Farrah2003, Farrah2013, Veilleux2002, Arribas2008, Bellocchi2013}. It is now well established that interactions and mergers are responsible for their intense star-forming activity specially at the higher end of the luminosity range \citep{Melnick1990, Clements1996, Bushouse2002}. The spectral energy distributions (SEDs) of U\slash LIRGs are dominated by dust thermal emission arising from reprocessing of ultraviolet radiation produced either by young massive stars or an active galactic nucleus (AGN) with dust temperatures in the range of 40--60\,K \citep{Symeonidis2010, Clements2018}. Local U\slash LIRGs have high star-formation efficiencies (SFEs;  defined as the ratio of the total infrared luminosity over the molecular gas mass) of the order of $>$100 $L_{\odot}\slash M_{\odot}$ \citep{Gao2004} and are characterized by compact star-forming regions, confined within the central kiloparsec of the merging systems \citep{Iono2009, Rujopakarn2011, Elbaz2011}. However, the star-forming regions are more spatially extended in isolated LIRGs and LIRGs in weakly interacting galaxy groups (see \citealt{AAH06s, Pereira2015not}).

At higher redshifts however, the properties of U\slash LIRGs appear to diverge from those of their local counterparts. Recent morphological studies have found that high-$z$ ULIRGs appear to be a mixture of merging\slash interacting systems and disk galaxies {(e.g., \citealt{Kartaltepe2012, Kaviraj2013, Wisnioski2018, Alcorn2018}).}
Continuous gas accretion via cold gas flows and minor mergers have been proposed as mechanisms to supply gas directly to the center of galaxies to maintain the high levels of star formation observed in these high-$z$ disk galaxies \citep{Ocvirk2008, Dekel2009}.
In addition, studies based on {\it Spitzer} and {\it Herschel} measurements have shown that the  infrared SEDs of high-$z$ U\slash LIRGs are more similar to those of local galaxies of lower luminosity, exhibiting stronger polycyclic aromatic hydrocarbon (PAH) features \citep{Pope2006, Farrah08} and colder dust temperatures (e.g., \citealt{Symeonidis2013, Muzzin2010,  Magdis2012}) than local U\slash LIRGs. 
\citet{Engelbracht08} suggested that such differences in the SEDs of local and high-$z$ U\slash LIRGs are qualitatively consistent with decreased metallicities by a factor of 1.5--2 while differences in the physical size of the star forming regions have also been implicated.

To fully understand the evolution of the ULIRG phenomenon, studies of the properties of U\slash LIRGs in the $0.2<z<1$ redshift range are necessary. Incidentally, this is the era when the Universe experienced a strong decrease in its SFR density \citep{Magnelli2011}. For this purpose, we have defined a sample of  intermediate-$z$ U\slash LIRGs originally drawn from {\it Herschel} observations of well-studied fields in the sky, including Bootes, Extended Groth Strip (EGS), the Cosmic Evolution Survey (COSMOS), and the {\it Hubble} Deep Field North (HDFN). A detailed study of the SEDs, the properties of their interstellar medium (ISM) and their molecular gas content have already been presented in \citet{Rigopoulou2014} and \citet{Magdis2014}. Spatially resolved kinematic studies of intermediate-$z$ U\slash LIRGs are a powerful diagnostic of the main source of dynamical support as they allow us to distinguish between relaxed virialized systems and merger events. 
{The morphology of lower-luminosity star-forming galaxies at intermediate redshifts show that only a few percent show signs of interaction (e.g., \citealt{Lee2017}). Similarly, the distance to the main-sequence seems to be enhanced by galaxy interactions. However, the star-formation efficiency seems to be independent of the interaction stage (e.g., \citealt{Bauermeister2013, Lee2017}).
}

In this paper, we present optical (H${\alpha}+$[\ion{N}{ii}] plus adjacent continuum) integral field spectroscopy (IFS) measurements of 9 systems observed with the SWIFT spectrograph \citep{ThatteSWIFT2006}. We focus on the study of the optical emission line spectra and kinematics probing the prevalence of AGN amongst the sample as well as the mechanism (e.g., merger vs. secular evolution) that drives star formation activity in U\slash LIRGs between the present day and $z\sim1$.

This paper is organized as follows: The SWIFT sample and observations are described in Section~\ref{s:obs}. The analysis of the H$\alpha$ and [\ion{N}{ii}] emission and the kinematic models are presented in Section~\ref{s:data}. In Section~\ref{s:discussion}, we compare the properties of our sample of  intermediate-$z$ U\slash LIRGs and local U\slash LIRGs. In Section \ref{s:conclusions}, we summarize the main results of this work.
Throughout this paper we assume the following cosmology: $H_{\rm 0} = 70$\,km\,s$^{-1}$\,Mpc$^{-1}$, $\Omega_{\rm m}=0.3$, and $\Omega_{\rm \Lambda}=0.7$.

\section{Observations and data reduction}\label{s:obs}

\subsection{The sample}

\begin{figure}
\centering
\includegraphics[width=0.42\textwidth]{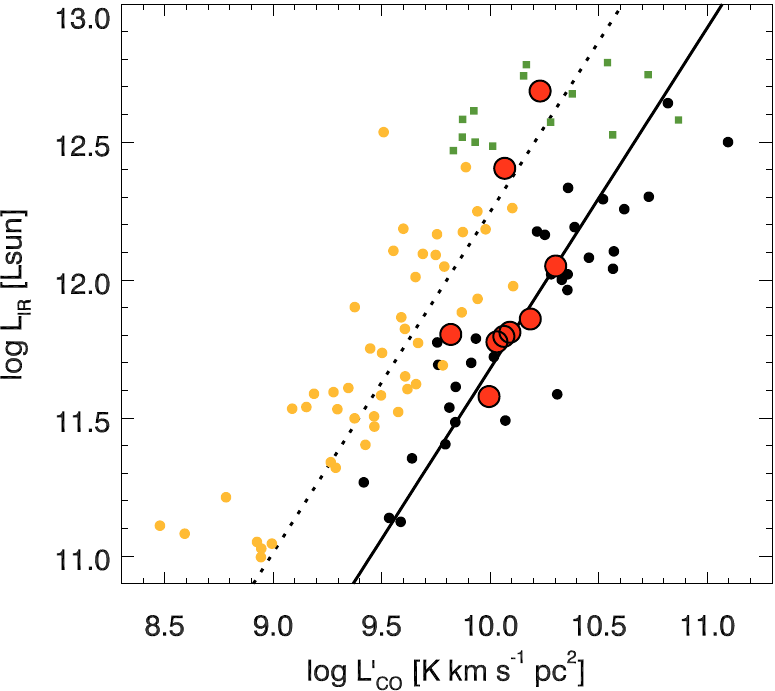}
\caption{\small { Infrared luminosity vs. CO(1-0) luminosity for the galaxies in the parent sample of intermediate-$z$ U\slash LIRGs with CO and [\ion{C}{ii}] observations (red circles; \citealt{Rigopoulou2014, Magdis2014}). Black circles are disk galaxies with $0.3<z<2.5$ \citep{Geach2011, Daddi2010b, Genzel2010}, orange circles are local U\slash LIRGs \citep{Solomon1997}, and green squares are warm $z\sim0.3$ U\slash LIRGs selected with {\it IRAS} \citep{Combes2011}. The solid and dashed black lines represent the relation observed for main sequence and starburst galaxies, respectively \citep{Sargent2014}.}}\label{fig:MS}
\end{figure}

We selected the targets from the sample of intermediate-$z$ U\slash LIRGs presented in \citet{Rigopoulou2014} and \citet{Magdis2014}.
In brief, { this sample of} intermediate-$z$ U\slash LIRGs were originally selected from the photometric catalogs from the {\it Herschel} Multi-tiered Extragalactic Survey (HerMES; \citealt{Oliver2012}) by applying the following two criteria: first, the sources had to have $S_{250}>$150 mJy and second, their redshift (photometric or spectroscopic) had to be $0.2<z<0.8$. The flux cut was imposed so that the sources would be IR luminous ($L_{\rm IR} >$ 10$^{11.5}L_{\odot}$) while the redshift cut { ensured} that the far-infrared line [\ion{C}{ii}]\,158$\mu$m line would shift in the wavelength range covered by the Spectral and Photometric Imaging REceiver-Fourier Transform Spectrometer (SPIRE-FTS, \citealt{Griffin2010SPIRE}) on board {\it Herschel}. 
No other criteria such as optical colors, morphologies, stellar mass, or presence of an AGN were taken into consideration. This is particularly important, as one of the goals of the current investigation is to establish whether intermediate-$z$ U\slash LIRGs are major mergers\slash interacting systems as is the case for most of their local counterparts. This initial selection resulted in a sample of 21 sources in the redshift range  $0.219<z< 0.887$. Amongst the targets there were two lensed sources XMM1 ($z_{\rm spec}$ = 2.308) and SWIRE6 ($z_{\rm spec}$ = 2.957), that were initially selected as the $z_{\rm spec}=0.502$ and $z_{\rm spec}=0.584$ foreground galaxies of the lensed SMGs HXMM01 \citep{Fu2013} and HLSW-01 \citep{Rigopoulou2018}.
All intermediate-$z$ U\slash LIRGs were subsequently targeted with the SPIRE-FTS. Of those, [\ion{C}{ii}] emission was detected in 17 of them. { Details of the spectroscopic CO follow up and SED modeling of this sample can be found in \citet{Magdis2014}}.

{ Most of the objects in the parent sample of intermediate-$z$ U\slash LIRGs are star-forming main sequence galaxies (see Figure~\ref{fig:MS} and \citealt{Magdis2014}). This figure shows that our sample has star-formation efficiencies (i.e., $L_{\rm IR}$ \slash $L^{\prime}_{\rm CO}$ ratio) similar to those of disk galaxies (black circles), and lie close to the relation observed in main-sequence galaxies (black solid line). For comparison, local U\slash LIRGs (orange circles) and intermediate-$z$ ULIRGs selected with {\it IRAS} (green squares), which are starburst objects, have higher star-formation efficiencies than most of the objects in our sample.}

{ For this paper, we selected the 9 northern intermediate-$z$ U\slash LIRGs for follow-up with the SWIFT integral field spectrograph.}
The SWIFT sample is presented in Table~\ref{tab:sample} with the naming convention as in \citet{Magdis2014}. 

\begin{table*}
\centering
\caption{Sample of intermediate-$z$ U\slash LIRGs}
\label{tab:sample}
\begin{tabular}{@{}lcccccccccc@{}}
\hline
Name & RA\,$^{a}$ & Dec.\,$^{a}$ & $z$\,$^{b}$ & ${\log L_{\rm IR}\slash L_\odot}$\,$^{c}$ & Separation\,$^{d}$ & Obs. date & Seeing FWHM\,$^{e}$ & {SFR$_{\rm IR}$}\,$^{f}$\\
& (J2000.0) & (J2000.0) &  &  & (arcsec\slash kpc) & & (arcsec\slash kpc)  & {(\Msun\,yr$^{-1}$)} \\
\hline
XMM2 W & 2:19:57.30 & -5:23:48.8 & 0.199 & 11.87 & 6.9\slash 23 & 2013-10-23 & 1.2\slash 3.9 & 110 \\
XMM2 E & 2:19:57.72 & -5:23:51.8 & 0.200 &  & \\
\hline
CDFS2 N & 3:28:18.02 & -27:43:07.5 & 0.248 & 11.83 & 6.8\slash 27 & 2013-10-24 & 1.5\slash 5.8 & 98 \\
CDFS2 S & 3:28:18.26 & -27:43:13.5 & 0.248 & \\
\hline
CDFS1 W & 3:29:04.39 & -28:47:53.0 & 0.289 & 11.80 & 7.0\slash 31 & 2013-10-25 & 1.4\slash 6.0 & 91 \\
CDFS1 E & 3:29:04.89 & -28:47:55.5 & 0.291 & \\
\hline
SWIRE5 W$^\star$ & 10:35:57.80 & +58:58:46.5 & 0.195 & \nodata & \nodata & 2013-01-16 & 2.7\slash 8.7 & \nodata \\
SWIRE5 E & 10:35:57.99 & +58:58:46.0 & 0.366 & 12.07 & & & 2.7\slash13.7 & 170 \\
\hline
SWIRE3 & 10:40:43.63 & +59:34:09.2 & 0.147 & 11.61 & \nodata & 2013-01-17 & 3.0\slash 7.7  & \nodata \\
\hline
SWIRE7 & 11:02:05.68 & +57:57:40.4 & 0.414 & 12.10 & \nodata & 2013-01-17 & 2.1\slash 11 & 190 \\
\hline
BOOTES1 & 14:36:31.95 & +34:38:29.2 & 0.351 & 12.69 & \nodata & 2012-05-12 & 1.5\slash 7.4  & \nodata \\
\hline
BOOTES2 & 14:32:34.90 & +33:28:32.2 & 0.249 & 11.90 & \nodata & 2012-05-14 & 1.4\slash 5.5 & 93 \\
\hline
FLS02 N & 17:13:31.49 & +58:58:04.4 & 0.436 & 12.42 & 3.6\slash 21 & 2012-05-09 & 1.6\slash 9.0 & 380 \\
FLS02 S & 17:13:31.64 & +58:58:01.0 & 0.437 &  &\\
\hline
\end{tabular}

\medskip
\raggedright \textbf{Notes:} 
$^{(a)}$ Coordinates from the $i$-band SDSS images except for the CDFS1 and CDFS2 systems for which we used the ALMA CO(3--2) emission (Rigopoulou et al. in prep.) and the R DSS image, respectively.
$^{(b)}$ Spectroscopic redshift from the SWIFT data presented here (see Tables~\ref{tab:fluxes} and \ref{tab:fluxes_broad}). 
$^{(c)}$ Total 8--1000\micron\ IR luminosity derived from the IR spectral energy distribution fitting \citep{Magdis2014}.
$^{(d)}$ Projected separation between the two nuclei of the system.
$^{(e)}$ Seeing FHWM measured from the telluric star observation and its equivalent linear size at distance of the observed system.
$^{(f)}$ {SFR derived from the total IR luminosity using the \citet{Murphy2011} calibration (see Section~\ref{ss:obscured}). We do not compute SFR$_{\rm IR}$ for SWIRE3 and BOOTES1 because they may host a bright AGN based on the broad H$\alpha$ emission line profile (Section~\ref{ss:integrated}) and this AGN might contribute to the observed IR luminosity.}
$^{(\star)}$ This is a foreground galaxy at $z=0.195$ unrelated to the ULIRG, SWIRE5 E, discussed in this paper. Note that the redshift of SWIRE5 W is coincident with the spectroscopic redshift given by \citet{RowanRobinson2010} for SWIRE5.
\end{table*}

\subsection{Optical integral field spectroscopy}

We obtained optical $I$- and $z$-band optical integral field spectroscopy of the U\slash LIRGs in our sample using the SWIFT instrument mounted on the 5.1\,m Hale Telescope at the Palomar Observatory in California. 
SWIFT covers the spectral range from 0.63 to 1.04\micron\ with a spectral resolving power $R$ of 3200 to 4400 ($\sim$90 to 70\,km\,s$^{-1}$). The data were obtained using the 0\farcs235 pixel scale which provides a field of view of 10\farcs3$\times$20\farcs9. The observations were carried out over four observing runs between 2012 and 2013. The median seeing measured from the telluric star observations is 1\farcs6 (see Table~\ref{tab:sample}).

We reduced the data using the SWIFT data reduction pipeline (see \citealt{Houghton2013}). This pipeline performs the bias subtraction, flat field correction, wavelength calibration, and produces a data cube for each frame. The observing strategy consisted of a nodding on integral field unit pattern which included the target in all the frames. Thanks to the relatively small size of these objects, it was possible to estimate the sky emission directly from the science frames. Therefore, the sky was subtracted using subsequent frames. The remaining hot-pixels\slash columns of the detector were manually masked. Finally, the sky-subtracted and masked data cubes for each frame were aligned and combined to obtain a final cube for each galaxy. The final cubes were corrected for telluric absorption and calibrated in flux using standard star observations taken at a similar air mass. We used the SDSS $i$- and $z$-band photometric observations available for 6 out of the 9 targets to assess the accuracy of the flux calibration. Calculating the synthetic photometry from the data cubes, we find that the SWIFT $i$- and $z$-band integrated fluxes are comparable to those of SDSS within $\pm$35\% on average. Therefore, we assume a 35\% uncertainty in the flux calibration of the SWIFT cubes.

\section{Data analysis}\label{s:data}

\subsection{Emission maps}\label{ss:maps}

We created H$\alpha$\,656.3\,nm and [\ion{N}{ii}]658.3\,nm emission line maps by modeling the spectrum of each spaxel of the data cubes. We fitted a model consisting of three Gaussian profiles with the same $\sigma$ to account for the H$\alpha$\,656.3\,mn, [\ion{N}{ii}]654.8\,nm, and [\ion{N}{ii}]658.3\,nm emission lines
and a linear function for the stellar continuum $\pm$5\,nm around these lines. The ratio between the fluxes of the two [\ion{N}{ii}] lines was fixed and set to [\ion{N}{ii}]654.8\,nm\slash [\ion{N}{ii}]658.3\,nm = 0.34 according to the expected theoretical ratio. {The H$\alpha$ stellar absorption is not taken into account because the SNR of stellar continuum is not high enough to model this absorption.}
We assumed a common line of sight velocity for the three lines, so the position of the three Gaussians is solely determined by this common line of sight velocity and their rest-frame wavelengths. The best-fit model was obtained by a $\chi^2$ minimization. In one of the systems, BOOTES1, we identified broad ($\sigma\sim$300--400\,km\,s$^{-1}$; FWHM$\sim750-900$\,km\,s$^{-1}$) and narrow ($\sigma<$100\,km\,s$^{-1}$) emission profiles in the three emission lines (see also next Section~\ref{ss:integrated}) in multiple spaxels. Therefore, we used a model with 6 Gaussian profiles (3 narrow and 3 broad) for this galaxy. The shift between the narrow and broad components was left free during the fit.

The H$\alpha$ and [\ion{N}{ii}]658.3\,nm line emission and continuum maps are shown in Figures ~\ref{fig:system_map_xmm2} and \ref{fig:map_xmm2w} and in Appendices~\ref{apx:emission_pairs} and \ref{apx:emission_indiv}. In Figure~\ref{fig:system_map_xmm2} and Appendix~\ref{apx:emission_pairs}, we show the continuum and line emission on the whole SWIFT field of view for the 4 close pair systems in our sample. The projected separation between the galaxies of these systems is 20--30\,kpc (Table~\ref{tab:sample}).
Figure ~\ref{fig:map_xmm2w} and Appendix~\ref{apx:emission_indiv} show the emission maps as well as the velocity field and the velocity dispersion ($\sigma$) for all the individual objects in our sample. 

For the kinematic analysis in Section~\ref{ss:kinematic}, we use the individual maps presented in Figure~\ref{fig:system_map_xmm2} and Appendix~\ref{apx:emission_pairs}. These emission line and kinematic maps have been clipped to a signal-to-noise ratio (SNR) $>$ 5 to exclude noisy spaxels from the modeling.

\begin{figure*}
\centering
\includegraphics[width=0.9\textwidth]{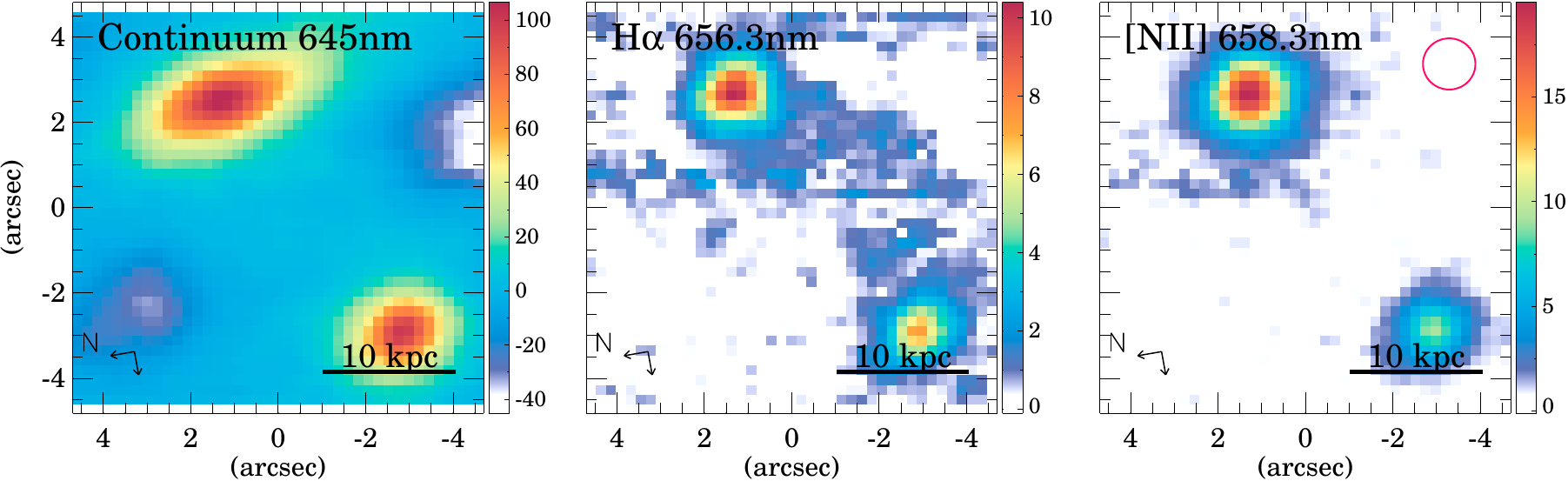}
\caption{\small SWIFT emission maps of the XMM2 system. The panels show: rest-frame 645\,nm continuum (left), H$\alpha$\,656.3\,mn (middle), and [\ion{N}{ii}]658.3\,nm (right). The units for the continuum and emission line maps are 10$^{-15}$\,erg\,cm$^{-2}$\,s$^{-1}$\,$\mu$m$^{-1}$ and 10$^{-17}$\,erg\,cm$^{-2}$\,s$^{-1}$, respectively. The scale represents 10 kpc at the distance of the system. The red circle in the right-hand panel corresponds to the seeing FWHM. The two arrows in the lower left corner indicate the North and East directions on the maps.
}\label{fig:system_map_xmm2}
\end{figure*}

\begin{figure*}
\centering
\includegraphics[width=0.8\textwidth]{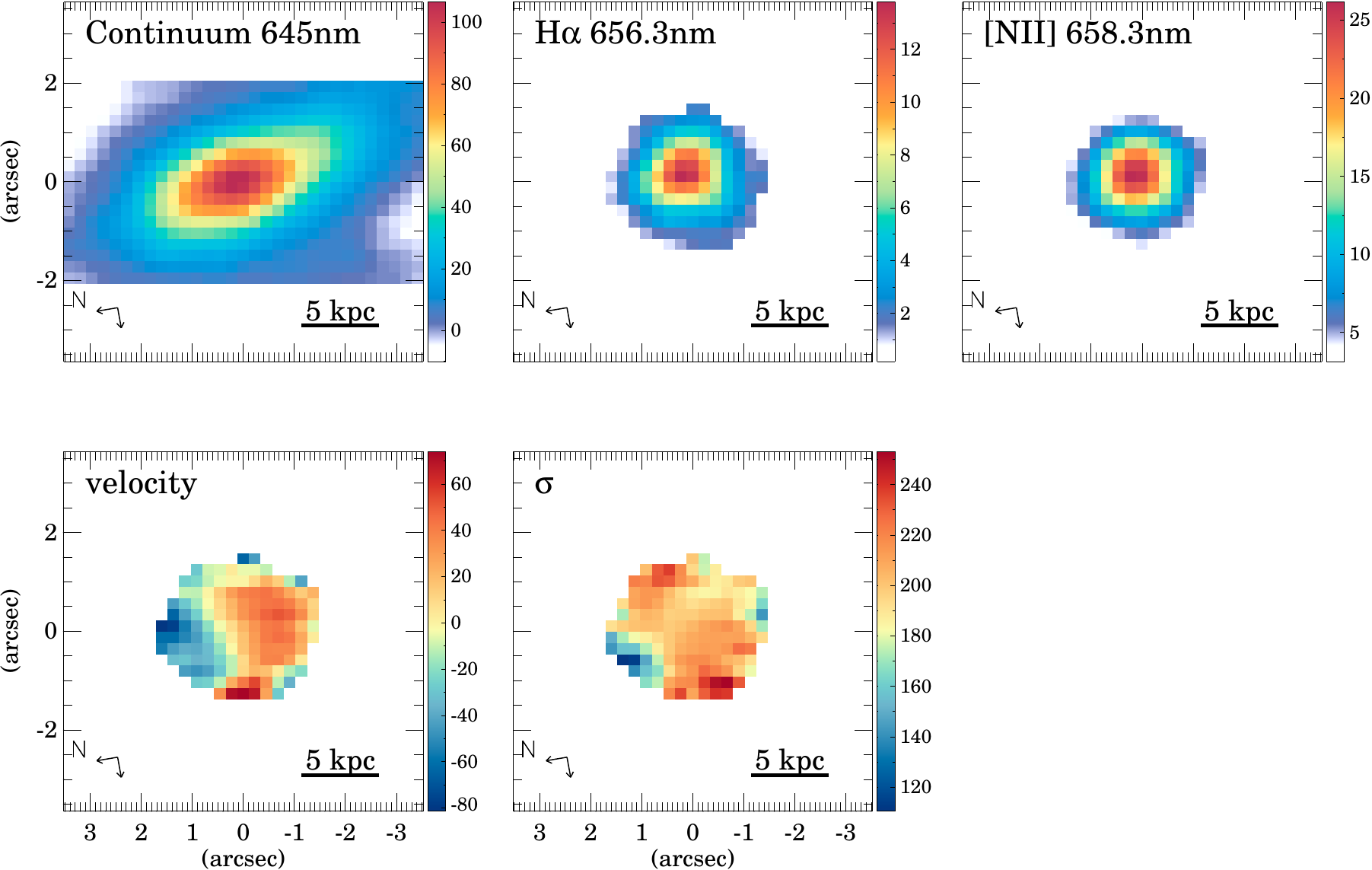}
\caption{\small SWIFT emission maps of XMM2 W. The top panels show: rest-frame 645\,nm continuum (left), H$\alpha$\,656.3\,mn (middle), and [\ion{N}{ii}]658.3\,nm (right). The units for the continuum and emission line maps are 10$^{-15}$\,erg\,cm$^{-2}$\,s$^{-1}$\,$\mu$m$^{-1}$ and 10$^{-17}$\,erg\,cm$^{-2}$\,s$^{-1}$, respectively. The bottom panels show the velocity field (left) with respect to the systemic velocity (see Table~\ref{tab:sample}) and the velocity dispersion $\sigma$ (right). Both velocity maps are in km\,s$^{-1}$.
The scale represents 5 kpc at the distance of the galaxy. The two arrows in the lower left corner indicate the North and East directions on the maps.
}\label{fig:map_xmm2w}
\end{figure*}

\subsection{Integrated spectra}\label{ss:integrated}

We extracted the integrated spectra of the individual objects in our sample by adding those spaxels where any emission line is detected at more than 3 times the standard deviation of the continuum. Similar to the spatially resolved analysis, we fitted a model consisting of a linear function for the stellar continuum and three Gaussian profiles with the same constraints described in Section~\ref{ss:maps}. For BOOTES1 and SWIRE3, we added 3 broad Gaussian profiles to account for the broad profiles observed in their integrated spectra. We note, that the SWIRE3 emission line is dominated by the broad component. Therefore, the narrow component is not detected in the individual spaxels and it is only apparent in the higher SNR integrated spectrum.

\begin{figure*}
\centering
\includegraphics[width=0.24\textwidth]{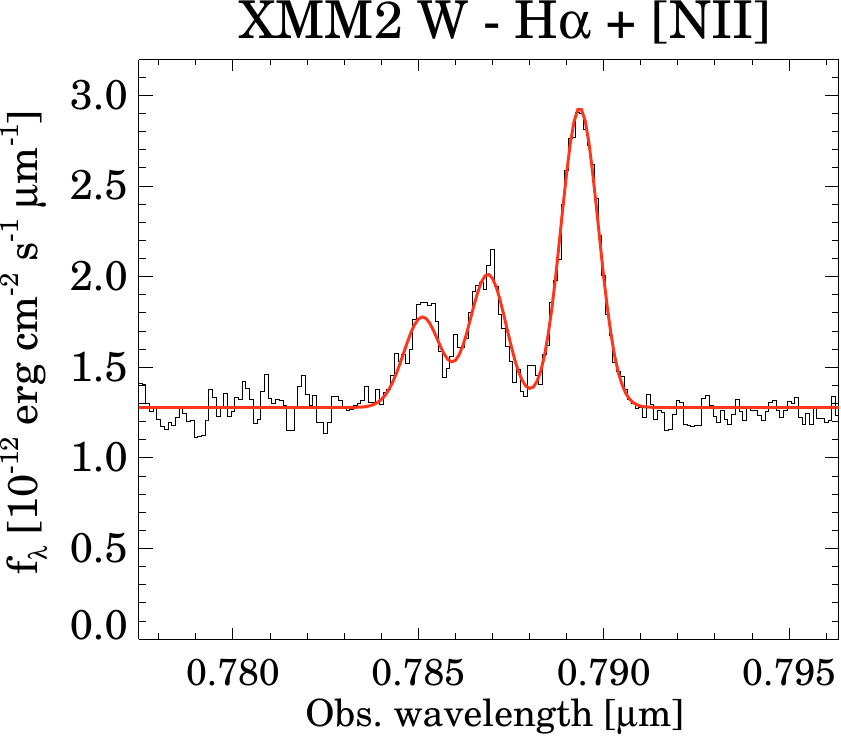}
\includegraphics[width=0.24\textwidth]{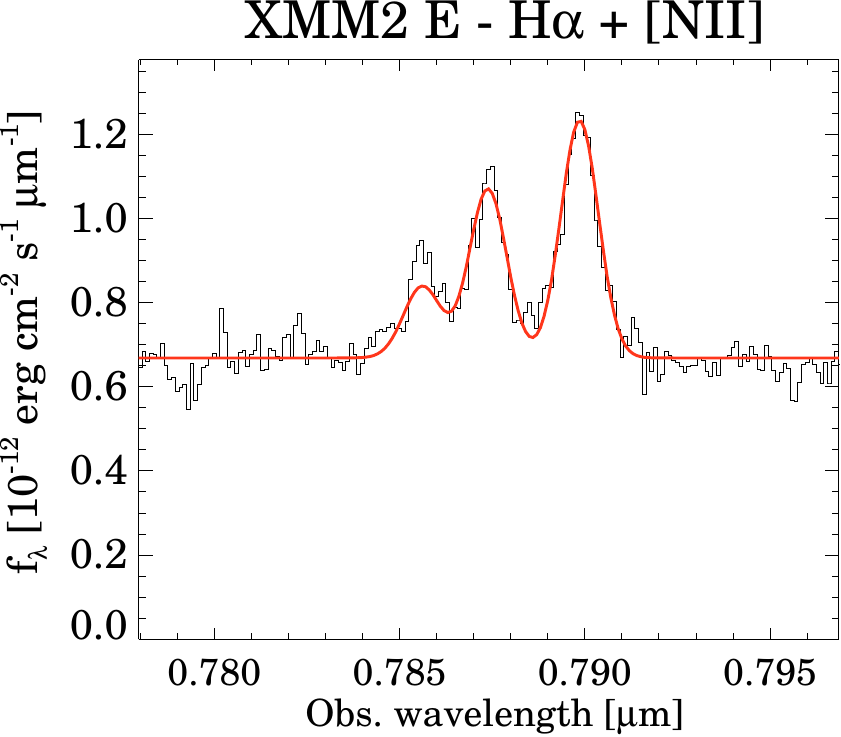}
\includegraphics[width=0.24\textwidth]{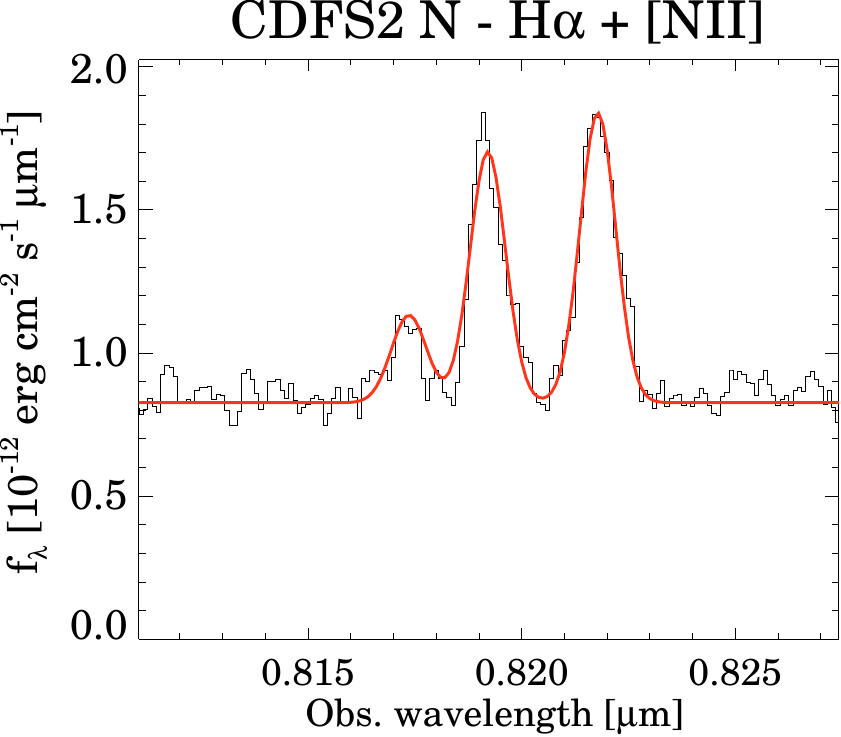}
\includegraphics[width=0.24\textwidth]{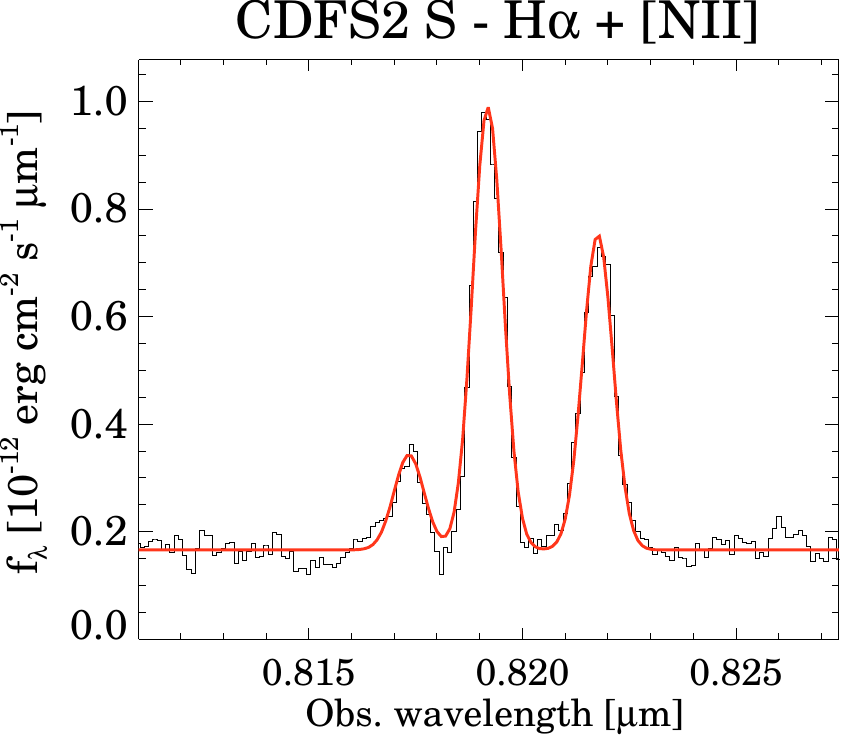}
\includegraphics[width=0.24\textwidth]{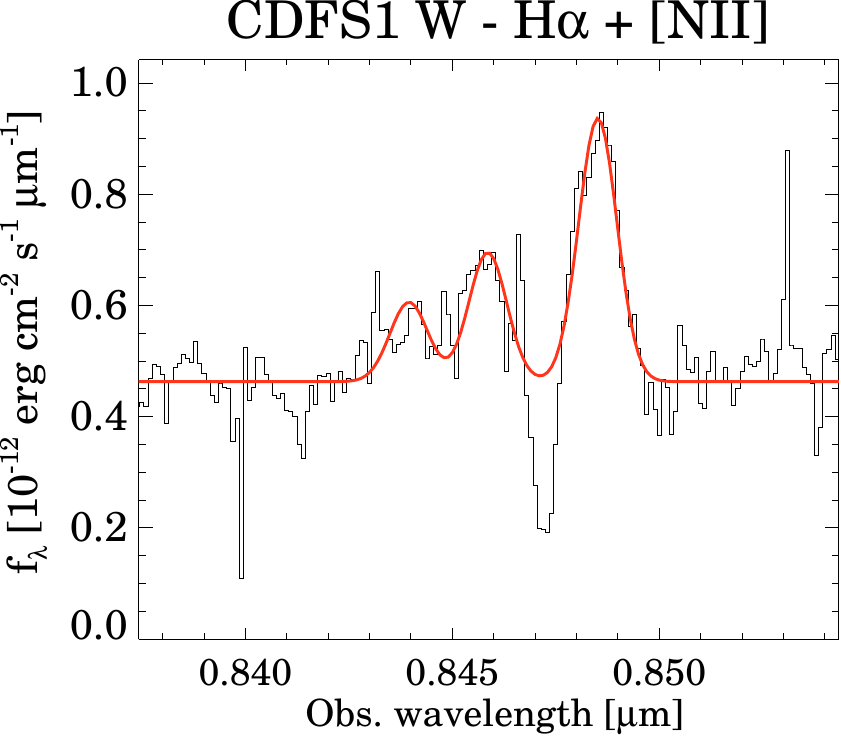}
\includegraphics[width=0.24\textwidth]{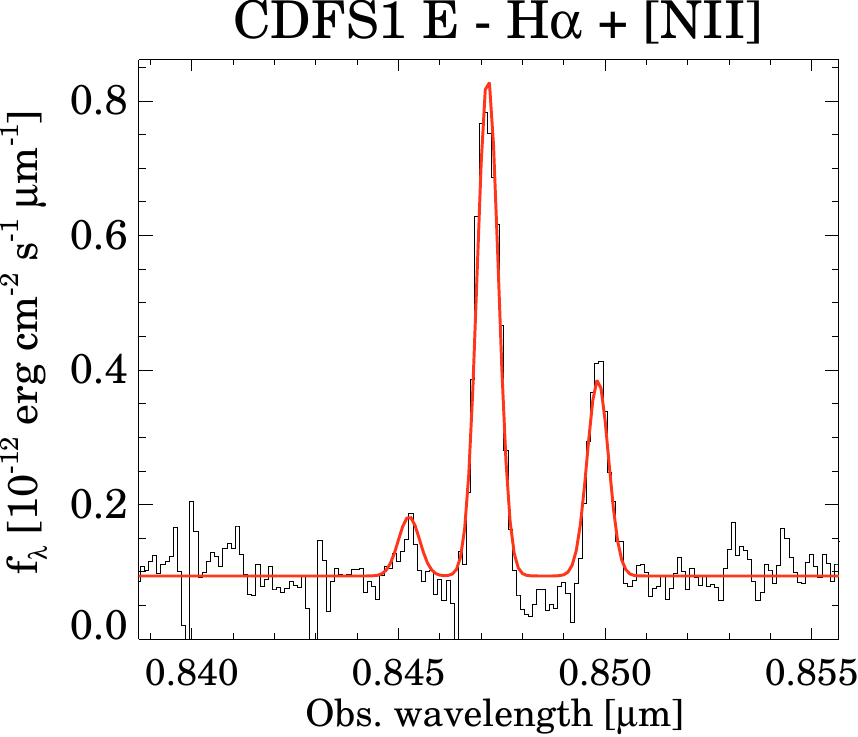}
\includegraphics[width=0.24\textwidth]{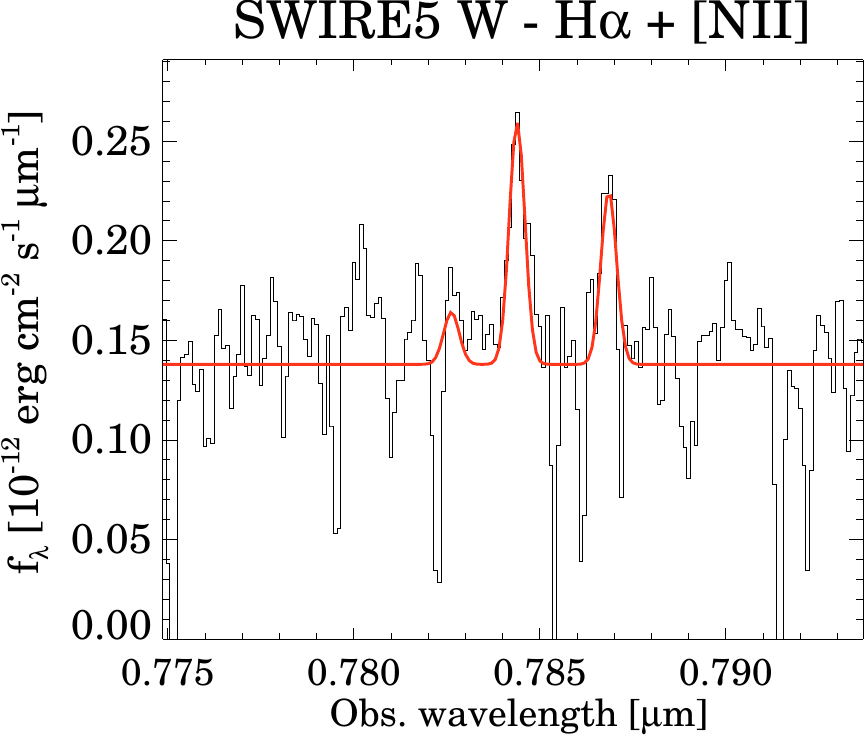}
\includegraphics[width=0.24\textwidth]{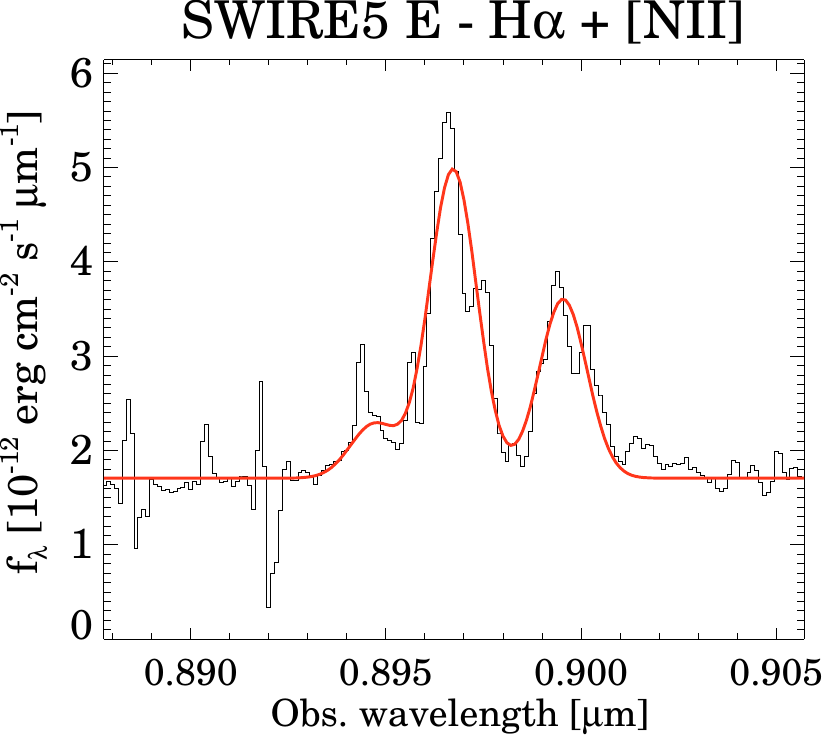}
\includegraphics[width=0.24\textwidth]{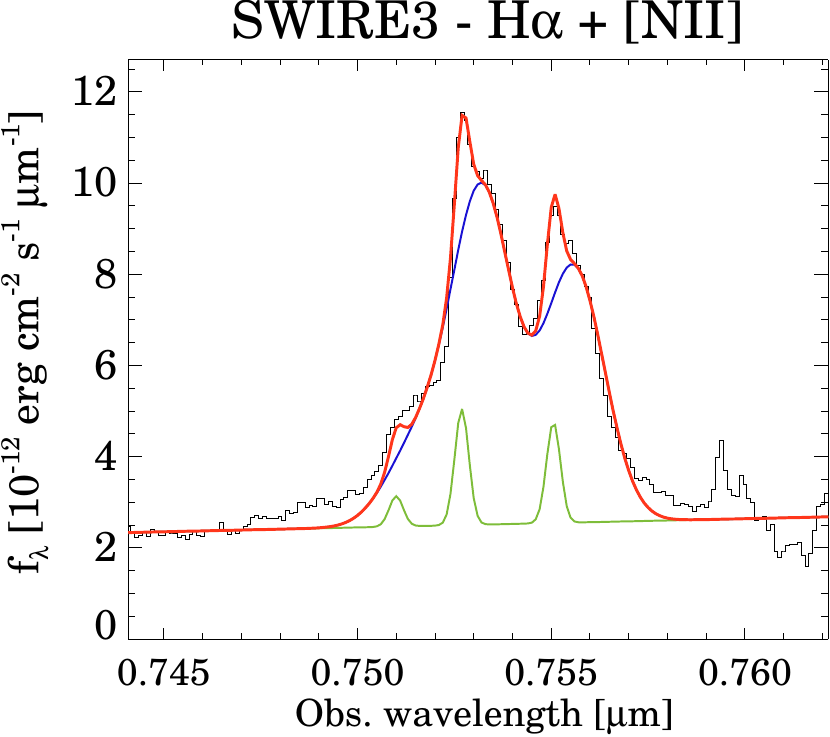}
\includegraphics[width=0.24\textwidth]{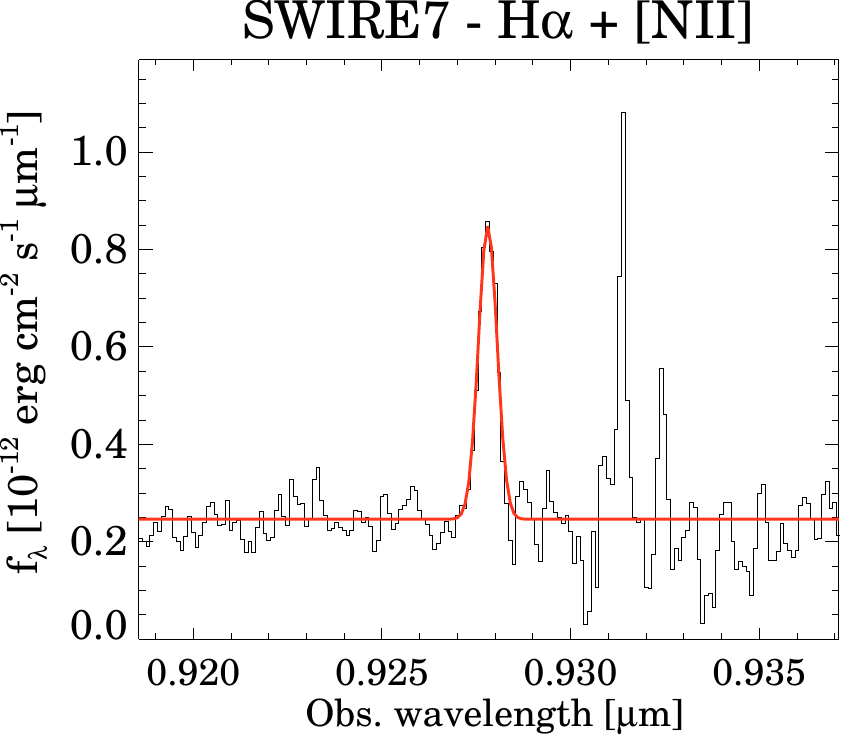}
\includegraphics[width=0.24\textwidth]{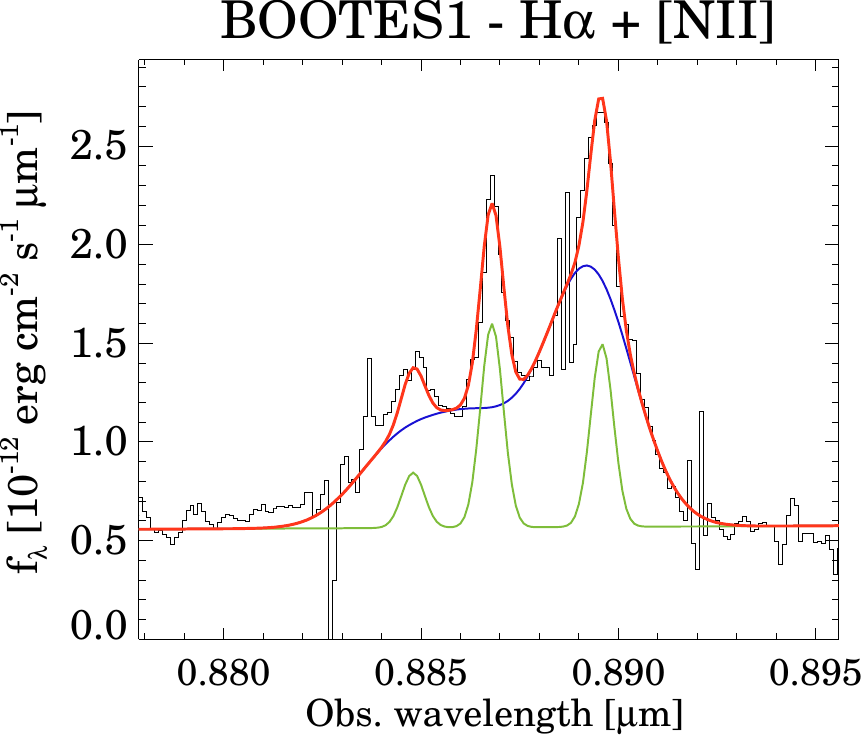}
\includegraphics[width=0.24\textwidth]{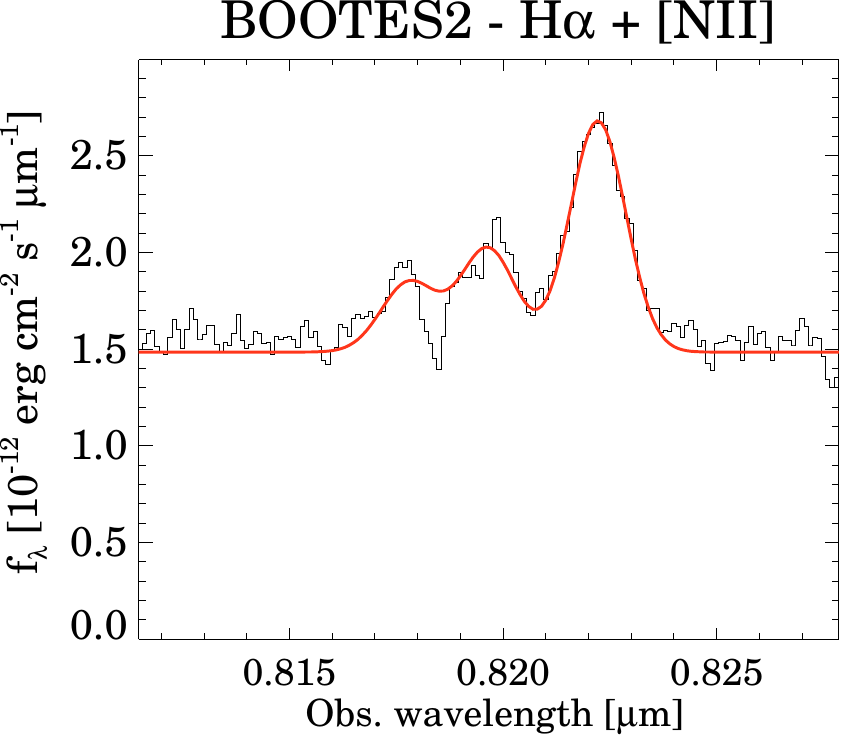}
\includegraphics[width=0.24\textwidth]{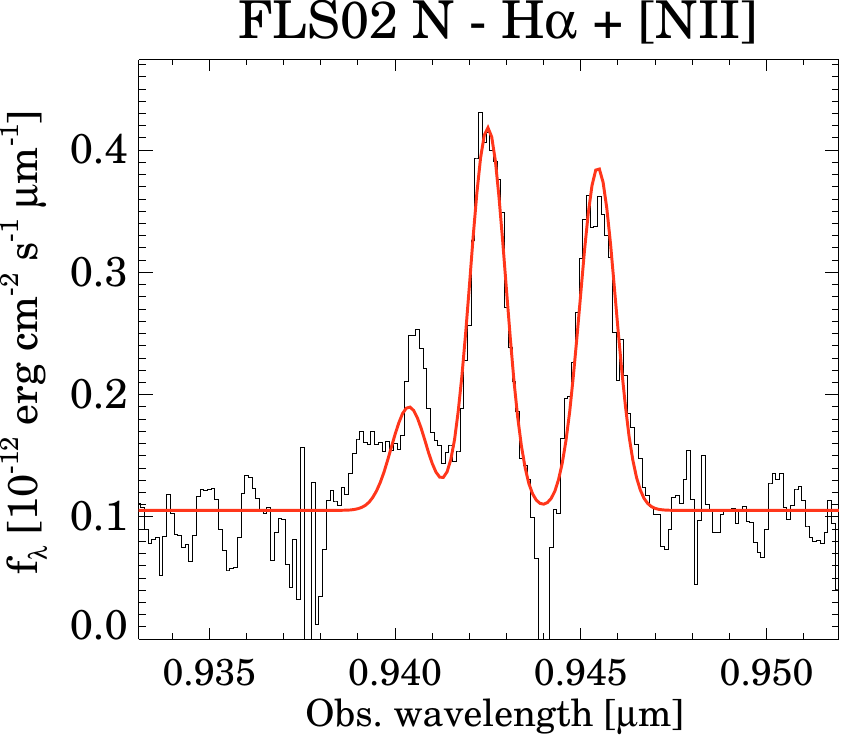}
\includegraphics[width=0.24\textwidth]{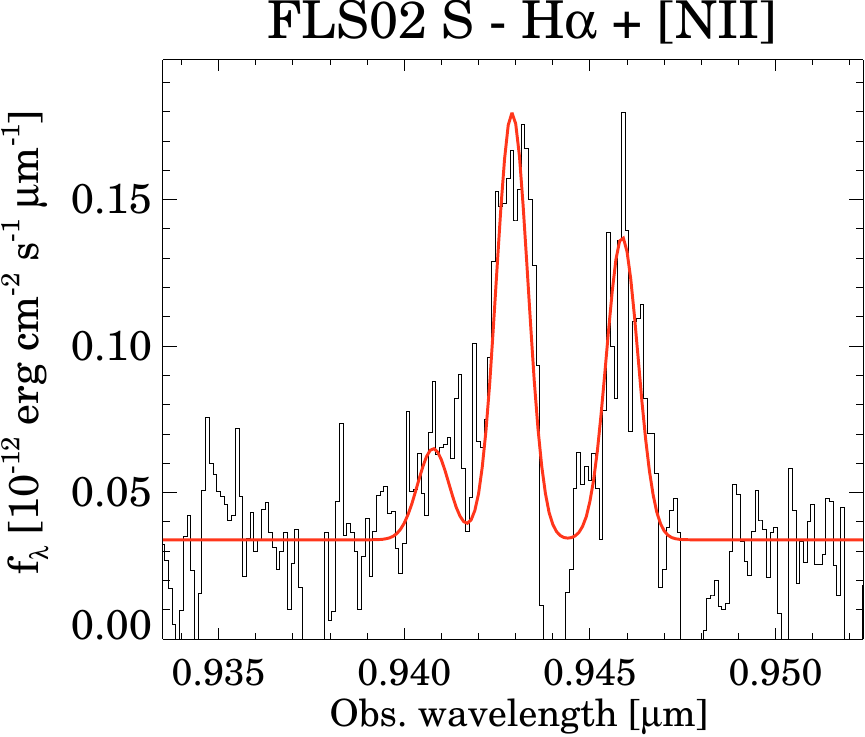}

\caption{\small SWIFT integrated spectra around the redshifted H$\alpha$ emission line for the individual objects (black histogram). The best-fit model is indicated by the red solid line. For BOOTES1 and SWIRE3, the best-fit model includes the narrow (green line) and broad (blue line) Gaussian profiles for the H$\alpha$ and [\ion{N}{ii}] emissions.
}\label{fig:integrated_sp}
\end{figure*}

The results of the fits are presented in Tables~\ref{tab:fluxes} and \ref{tab:fluxes_broad} for the galaxies with narrow and narrow plus broad line profiles, respectively. We show the observed spectra and the best-fit models in Figure~\ref{fig:integrated_sp}.

\begin{table*}
\centering
\caption{H$\alpha$ and [\ion{N}{ii}]658.3\,nm emission properties.}
\label{tab:fluxes}
\begin{tabular}{@{}lcccccccccccc@{}}
\hline
Name & ${\lambda_{\rm H\alpha}}^{a}$ & F(H$\alpha$)$^{b}$ & $\frac{\rm [NII]658.3\,nm}{\rm H\alpha}$\,$^{c}$ & v$^{d}$ & $\sigma$\,$^{e}$ & $L$(H$\alpha$)$^{f}$ & SFR(H$\alpha$)$^{g}$ \\
& (nm) & (10$^{-16}$\,erg\,cm$^{-2}$\,s$^{-1}$) & & (km\,s$^{-1}$) & (km\,s$^{-1}$) & (10$^{40}$\,erg\,s$^{-1}$) & ($M_\odot$\,yr$^{-1}$)\\
\hline
XMM2 W & 786.9 $\pm$ 0.9 & 9.22 $\pm$ 0.33 & 2.26 $\pm$ 0.09 & 53817 $\pm$ 63 & 191 $\pm$ 53 & 10.5 & 0.56 \\
XMM2 E & 787.4 $\pm$ 1.3 & 5.10 $\pm$ 0.17 & 1.41 $\pm$ 0.06 & 54004 $\pm$ 91 & 192 $\pm$ 78 & 5.8 & 0.31 \\
CDFS2 N & 819.2 $\pm$ 0.6 & 9.15 $\pm$ 0.21 & 1.16 $\pm$ 0.03 & 65409 $\pm$ 48 & 159 $\pm$ 71 & 17.1 & 0.92 \\
CDFS2 S & 819.2 $\pm$ 0.2 & 7.29 $\pm$ 0.04 & 0.71 $\pm$ 0.01 & 65402 $\pm$ 12 & 129 $\pm$ 50 & 13.6 & 0.73 \\
CDFS1 W & 845.9 $\pm$ 0.8 & 2.68 $\pm$ 0.12 & 2.05 $\pm$ 0.10 & 74483 $\pm$ 69 & 163 $\pm$ 51 & 7.1 & 0.38 \\
CDFS1 E & 847.2 $\pm$ 0.1 & 4.79 $\pm$ 0.02 & 0.39 $\pm$ 0.01 & 74915 $\pm$ 6 & 90 $\pm$ 11 & 12.9 & 0.69 \\
SWIRE5 W$^\star$ & 784.4 $\pm$ 0.9 & 0.64 $\pm$ 0.03 & 0.72 $\pm$ 0.05 & 52896 $\pm$ 60 & 80 $\pm$ 2 & 0.7 & 0.04 \\
SWIRE5 E & 896.7 $\pm$ 0.4 & 50.13 $\pm$ 0.48 & 0.58 $\pm$ 0.01 & 90655 $\pm$ 39 & 204 $\pm$ 37 & 230.2 & 12.36 \\
SWIRE7 & 927.8 $\pm$ 0.2 & 3.73 $\pm$ 0.05 & \nodata & 99839 $\pm$ 19 & 80 $\pm$ 2 & 22.8 & 1.23 \\
BOOTES2 & 819.7 $\pm$ 1.8 & 8.74 $\pm$ 0.45 & 2.24 $\pm$ 0.13 & 65560 $\pm$ 140 & 237 $\pm$ 98 & 16.4 & 0.88 \\
FLS02 N & 942.5 $\pm$ 0.2 & 3.76 $\pm$ 0.03 & 0.90 $\pm$ 0.01 & 104004 $\pm$ 17 & 152 $\pm$ 74 & 26.1 & 1.40 \\
FLS02 S & 942.9 $\pm$ 0.1 & 1.55 $\pm$ 0.01 & 0.71 $\pm$ 0.01 & 104118 $\pm$ 15 & 135 $\pm$ 56 & 10.8 & 0.58 \\
\hline
\end{tabular}

\medskip
\raggedright \textbf{Notes:} 
$^{(a)}$ Observed wavelength of the H$\alpha$ transition. 
$^{(b)}$ Observed H$\alpha$ flux. 
$^{(c)}$ Observed [NII]658.3\,nm\slash H$\alpha$ ratio. 
$^{(d)}$ Velocity derived from the H$\alpha$ and [NII]658.3\,nm emission using the relativistic velocity definition. 
$^{(e)}$ Width of the best-fit Gaussian profiles. 
$^{(f)}$ Observed H$\alpha$ luminosity. 
$^{(g)}$ SFR derived from the observed H$\alpha$ luminosity using the \citet{Murphy2011} calibration.
$^{(\star)}$ SWIRE5 W is a foreground galaxy not related to the ULIRG SWIRE5 E (see Table~\ref{tab:sample}).
\end{table*}

\subsection{Kinematic models}\label{ss:kinematic}

To analyze the kinematic information in the data cubes we used GalPak$^{\rm 3D}$ \citep{Bouche2015}. This software takes the observed data cube as input and, using a Bayesian approach, produces a best-fit rotating disk model. To determine the best-fit parameters, GalPak$^{\rm 3D}$ convolves the model with both the point spread function (Table~\ref{tab:sample}) and the spectral line spread function (about 80\,km\,s$^{-1}$). Doing so, it is possible to recover the intrinsic velocity dispersion and amplitude of the rotation curve even when the galaxies are just marginally spatially resolved like in our case.

For the rotating disk model, we assume that the H$\alpha$ emission intensity in these galaxies follows a 2D Gaussian distribution, that the rotation curve can be modeled with an arctan functional form, and that the intrinsic velocity dispersion ($\sigma$) is spatially constant in the disk (see \citealt{Bouche2015} for details).
Then, GalPak$^{\rm 3D}$ finds the spatial parameters (radius, inclination, and position angle) for the 2D Gaussian intensity distribution as well as the kinematic parameters (intrinsic velocity dispersion, systemic velocity, turnover radius, and maximum velocity) for the arctan model.

We show the best-fit models in Figure \ref{fig:galpak_xmm2w} and in Appendix~\ref{apx:models}. These models can reproduce the observed flux with residuals lower than 10--20\% for most of the galaxies. For the velocity fields, the typical residual is lower than 20\,km\,s$^{-1}$. For the $\sigma$ maps, the residuals are also about 20\,km\,s$^{-1}$. The best-fit parameters are listed in Table~\ref{tab:galpak}.

\begin{table*}
\centering
\caption{Parameters of the best-fit GalPak$^{\rm 3D}$ model}
\label{tab:galpak}
\begin{tabular}{@{}lcccccccccccc@{}}
\hline
Name & $i$\,$^{a}$ & PA\,$^{b}$ & R\,$^{c}$ & v$_{\rm max}$\,$^{d}$ & $\sigma$\,$^{e}$ & v\slash$\sigma$\,$^{f}$ & $\frac{f_{\rm gas}({\rm z = 0)}}{f_{\rm gas}({\rm z = z_{\rm U\slash LIRG}})}$~$^{g}$ & $\sigma_{\rm z_0}$~$^{h}$ \\
& (deg) & (deg) & (arcsec\slash kpc) & (km\,s$^{-1}$) & (km\,s$^{-1}$) & & & (km\,s$^{-1}$)\\
\hline
XMM2 W   & 62 & 5   & 0.66\slash2.2 &  86 $\pm$ 9 & 195 $\pm$ 10 & 0.44 $\pm$ 0.05 & 0.73 & 142 \\
XMM2 E   & 35 & 26  & 0.41\slash1.4 & 119 $\pm$10 & 212 $\pm$ 15 & 0.56 $\pm$ 0.06 & 0.73 & 155 \\
CDFS2 S  & 46 & -48 & 0.61\slash2.4 &  78 $\pm$16 & 109 $\pm$ 8  & 0.72 $\pm$ 0.16 & 0.69 & 75 \\
CDFS1 W  & 67 & 125 & 2.58\slash11 & 265 $\pm$ 9 &  84 $\pm$ 5  & 3.15 $\pm$ 0.22 & 0.65 & 55 \\
CDFS1 E  & 48 & -49 & 0.82\slash3.6 & 115 $\pm$ 6 &  74 $\pm$ 6  & 1.55 $\pm$ 0.15 & 0.65 & 48 \\
SWIRE5 E & 37 & 19  & 0.92\slash4.7 & 310 $\pm$11 & 114 $\pm$ 9  & 2.72 $\pm$ 0.24 & 0.60 & 68 \\
SWIRE7   & 30 & 51  & 0.48\slash2.7 & 144 $\pm$18 &  52 $\pm$ 8  & 2.77 $\pm$ 0.55 & 0.53 & 76 \\
BOOTES2  & 44 & 47  & 0.80\slash3.1 & 108 $\pm$ 7 & 215 $\pm$ 12 & 0.50 $\pm$ 0.04 & 0.69 & 148 \\
\hline
\end{tabular}
\medskip
\raggedright \textbf{Notes:} 
$^{(a)}$ Galaxy inclination.
$^{(b)}$ Position angle (East of North) of the blue-shifted part of the kinematic major axis.
$^{(c)}$ FWHM\slash 2 of the 2D Gaussian spatial model in angular and linear units at the distance of the each object.
$^{(d)}$ De-projected semi-amplitude of the rotation curve.
$^{(e)}$ Intrinsic velocity dispersion of the disks. This model assumes that the velocity dispersion is uniform within the disk.
$^{(f)}$ Dynamical ratio between the de-projected velocity semi-amplitude and the intrinsic velocity dispersion. 
$^{(g)}$ Ratio between the gas fractions at $z=0$ and the redshift of each target used to scale the observed $\sigma$ in Figure~\ref{fig:vsigma} (see Section~\ref{ss:dyn_ratio} and Equation~\ref{eqn:sigma} for details).
$^{(h)}$ Equivalent velocity dispersion at $z=0$ derived using Equation~\ref{eqn:sigma} and ${f_{\rm gas}({\rm z = 0)}}\slash {f_{\rm gas}({\rm z = z_{\rm U\slash LIRG}})}$ (see Section~\ref{ss:dyn_ratio}).
\end{table*}

We excluded from the kinematic modeling CDFS2\,N, SWIRE3, BOOTES1, FLS02\,N, and FLS02\,S, so in total we have modeled 8 objects with GalPak$^{\rm 3D}$. For CDFS2\,N, the emission map and the velocity field might be compatible with two counter-rotating interacting galaxies approximately along the East-West direction separated by $\sim$10\,kpc (see Figure~\ref{fig:map_cdfs2n}). SWIRE3 and BOOTES1 have broad H$\alpha$ emission (see Figure~\ref{fig:integrated_sp}) so the narrow emission, associated with the disk rotation, is not well constrained.
For FLS02\,N, the H$\alpha$ emission seems to be spatially unresolved so no kinematic information is available from this transition (see Figure~\ref{fig:map_fls02n}), and FLS02\,S contains too few spaxels with H$\alpha$ emission for a reliable modeling (see Figure~\ref{fig:map_fls02s}). 

{ SDSS images are available for 7 out of 9 systems in our sample. 
However, the limited angular resolution and sensitivity of these images is not sufficient to study the morphology of these objects, identify perturbed morphologies, or detect tidal tails.
High angular resolution imaging of these system would be needed to firmly establish their actual morphology.}

\begin{figure*}
\centering
\includegraphics[width=0.8\textwidth]{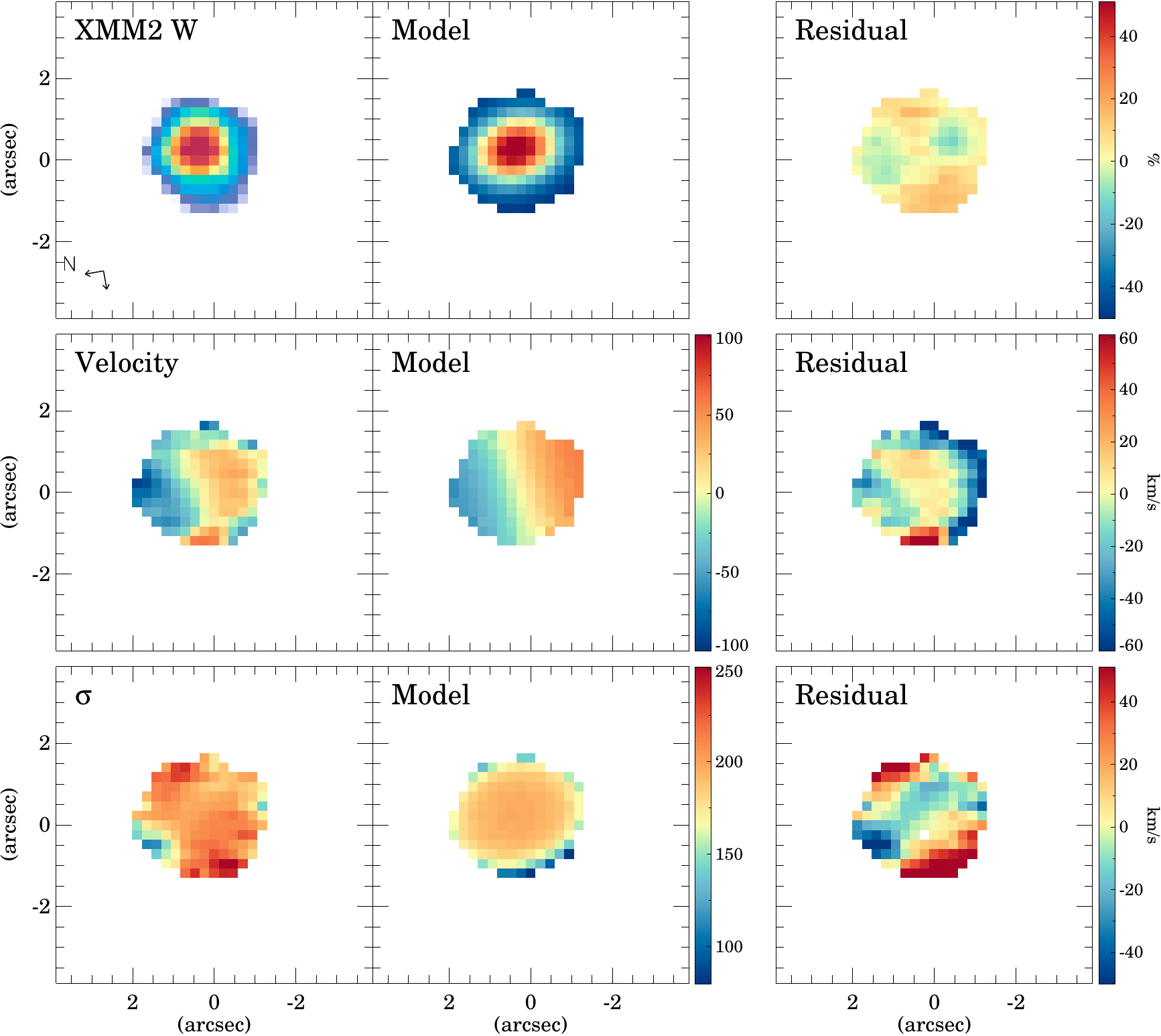}
\caption{\small GalPak$^{\rm 3D}$ model of XMM2 W. The first column shows the observed flux, velocity, and velocity dispersion ($\sigma$). The second column is the best-fit model, and the third column the residuals. The velocity and $\sigma$ panels are in km\,s$^{-1}$ units. The residual flux map shows the difference between the observed and the model in percentage. The two arrows in the lower left corner indicate the North and East directions on the maps.
}\label{fig:galpak_xmm2w}
\end{figure*}

\section{Discussion}\label{s:discussion}

\citet{Magdis2014} found that intermediate-$z$ ($0.2 <z<0.9$) U\slash LIRGs have physical properties that differ from those of local U\slash LIRGs. Based on \textit{Herschel} [\ion{C}{ii}]158\micron\ and ground-based CO observations, these intermediate-$z$ U\slash LIRGs present stronger [\ion{C}{ii}] emission with respect to the total $L_{\rm IR}$ and lower star-formation efficiencies than local ULIRGs. To explain these differences, \citet{Magdis2014} proposed that the starbursts of these intermediate-$z$ U\slash LIRGs are not driven by galaxy mergers and that, unlike local ULIRGs, they might take place in extended regions within the galaxy. Also, \citet{Magdis2014} suggested that the evolution of the metal content of the U\slash LIRGs at these intermediate-$z$  might also explain the observed physical properties.

To investigate these possibilities, in this section, we compare the kinematics, SFR obscuration, metal abundances, and dust temperatures of local U\slash LIRGs and our sample of intermediate-$z$ U\slash LIRGs.

\subsection{Dynamical ratio}\label{ss:dyn_ratio}

The dynamical ratio defined as the ratio between the velocity amplitude of the rotation curve and the velocity dispersion can be used to parametrize the kinematic state of a system (e.g., \citealt{Epinat2010, Bellocchi2013}). Objects that are rotationally supported (i.e., disks) have high $v\slash \sigma$ ratios, whereas interacting and merging systems have higher velocity dispersions ($\sigma$) and, therefore, lower $v\slash \sigma$ ratios.

Assuming that disks are thin and quasi-stable, the velocity dispersion is expected to increase with $z$ due to the increased gas fractions ($f_{\rm gas}$) at higher $z$ (see \citealt{Wisnioski2015, Krumholz2018}). To account for this intrinsic increase of the velocity dispersion, we apply the following corrections to the observed  $\sigma$ based on \citet{Wisnioski2015} and obtain an equivalent velocity dispersion at $z=0$ ($\sigma_{\rm z_0}$). Using their equations 3 to 6, we can determine the evolution of $f_{\rm gas}$ with $z$ combining the evolutions of the depletion time and specific SFR (see \citealt{Wisnioski2015}). Then, the obtained $f_{\rm gas}(z)$ is proportional to $\sigma(z)$ based on the Toomre stability criterion (see equation 8 of \citealt{Wisnioski2015}).
Finally, taking the ratio between the velocity dispersion at $z=0$ and that at the redshift of each U\slash LIRG, we obtain the following relation:
\begin{equation}
\sigma_{\rm z_0}\slash \sigma_{\rm z_{\rm U\slash LIRG}} = f_{\rm gas}({\rm z = 0)} \slash f_{\rm gas}({\rm z = z_{\rm U\slash LIRG})}
\label{eqn:sigma}
\end{equation}
where $\sigma_{\rm z_{\rm U\slash LIRG}}$ is the observed velocity dispersion in the intermediate-$z$ U\slash LIRGs, and $\sigma_{\rm z_0}$ the equivalent velocity dispersion at $z=0$. Using this $\sigma_{\rm z_0}$, we obtain dynamical ratios that can be directly compared to those measured in local galaxies.

For this comparison, we plot the dynamical ratio for the 8 intermediate-$z$ U\slash LIRGs whose gas kinematics were modeled with GalPak$^{\rm 3D}$ (see Section~\ref{ss:kinematic}) using the corrected $\sigma_{\rm z_0}$ values. {Galaxy pairs are considered individually in this Section.}
We also plot the dynamical ratios of local U\slash LIRGs with similar optical IFS data from \citet{GarciaMarin2009Part1} and \citet{Bellocchi2013}. We limit the local sample to galaxies with $\log L_{\rm IR}\slash L_\odot>11.6$ to match the luminosity range of the intermediate-$z$ objects. 
From the dynamical ratios and velocity dispersions observed in isolated U\slash LIRGs (defined as class 0 objects in \citealt{Bellocchi2013}), we determine a threshold value to separate isolated and interacting galaxies as the mean value of isolated objects plus (minus) three times the standard deviation of the velocity dispersion (dynamical ratio). These threshold values are $\sigma<56$\,km\,s$^{-1}$ and v$\slash \sigma>3.2$. {We also compare with the velocity dispersions and dynamical ratios of local elliptical galaxies from \citealt{Cappellari2007}. Similarly the threshold between interacting and elliptical is calculated using the mean dynamical ratio plus \slash minus three times its standard deviation. We note that dynamical ratios between 0.3 and 0.8 are observed in both elliptical and interacting systems.}

Using the criteria described above, we find that half, 4 out of 8, of the SWIFT intermediate-$z$ U\slash LIRGs are compatible with the kinematic parameters measured in interacting systems {and elliptical galaxies}. 
Three of these objects belong to systems with various galaxies (CDFS2 S\footnote{CDFS2 N is not modeled with GalPak$^{\rm 3D}$ since it shows a complex kinematics which does not resemble a rotating disk (see Section~\ref{ss:kinematic}).}, XMM2 E and W). For the fourth object, BOOTES2, the continuum image suggests that it is an isolated object. This might indicate a more advanced phase of the interaction where the separation between the 2 nuclei is not resolved by the SWIFT data (5.5\,kpc spatial resolution for this object).
CDFS1 W and SWIRE7 lie in the region of isolated systems, so they might actually be isolated galaxies (SWIRE7) or be at an early stage of the interaction (CDFS1 W). Finally, two objects (SWIRE5 E and CDFS1 E) have intermediate dynamical ratios and velocity dispersions.
This diagram suggests that some fraction (1--2 out of 8) of the intermediate-$z$ U\slash LIRGs ($0.2<z<0.4$) are isolated disks.

\begin{figure}
\centering
\includegraphics[width=0.4\textwidth]{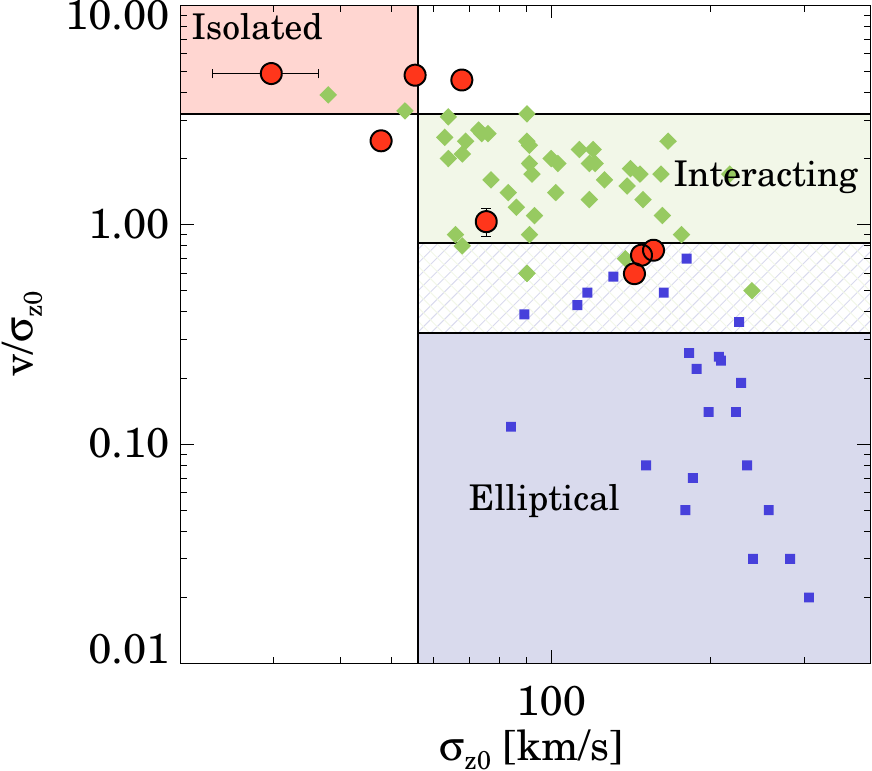}
\caption{\small {Ratio between the amplitude of the rotation curve and the velocity dispersion for the SWIFT intermediate-$z$ U\slash LIRGs (red circles), local U\slash LIRGs with $\log L_{\rm IR}\slash L_\odot> 11.6$ (green diamonds; \citealt{GarciaMarin2009Part1, Bellocchi2013}), and local ellipticals (blue squares; \citealt{Cappellari2007}). The observed velocity dispersions of the intermediate-$z$ U\slash LIRGs have been corrected to account for the expected increase of the velocity dispersion with $z$ (see Section~\ref{ss:dyn_ratio}).
The red, green, and blue  shaded areas indicate the typical ratios observed in isolated, interacting, and elliptical objects. The area filled with green and blue lines indicates a range a ratios common for both interacting and elliptical galaxies (see Section~\ref{ss:dyn_ratio}).}}\label{fig:vsigma}
\end{figure}

\subsection{Obscured star-formation}\label{ss:obscured}

Star-formation in local interacting U\slash LIRG usually occurs in compact ($r<1$\,kpc) nuclear regions (e.g., \citealt{Rujopakarn2011, Arribas2012}). 
These compact starbursts are heavily affected by dust extinction (e.g., \citealt{Rigopoulou1999, Piqueras2013}) and the SFRs derived from the optical H$\alpha$ emission can be a few orders of magnitude lower than those derived from IR tracers (e.g., \citealt{Hopkins2001, Dopita2002, RodriguezZaurin2011}).

By contrast, higher redshift ($z>1-3$) U\slash LIRGs have more extended SF ($r>2$\,kpc; e.g., \citealt{Iono2009, Simpson2015, Rujopakarn2016}) and are less affected by dust extinction (e.g., \citealt{Buat2015}).

\begin{figure}
\centering
\includegraphics[width=0.4\textwidth]{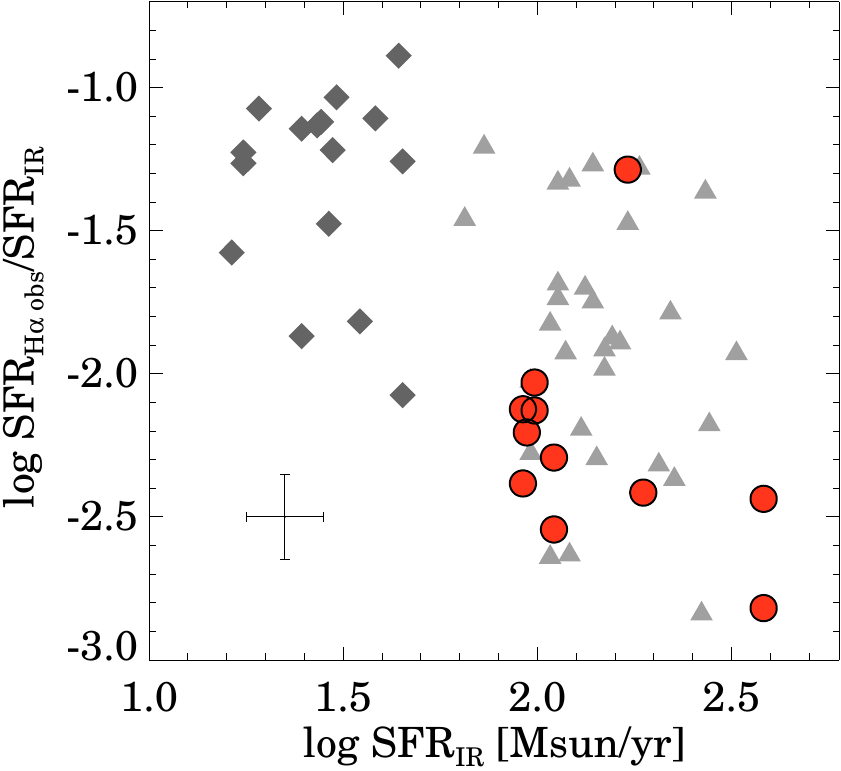}
\caption{\small Ratio between the SFR from the observed H$\alpha$ emission (not corrected for extinction) and the IR SFR as a function of the IR SFR. The red circles correspond to the SWIFT intermediate-$z$ U\slash LIRGs and the gray triangles to local U\slash LIRGs ($\log L_{\rm IR}\slash L_\odot > 11.6$) from \citet{RodriguezZaurin2011}. The dark gray diamonds correspond to lower luminosity LIRGs also from \citet{RodriguezZaurin2011}. The average uncertainties are indicated by the error bars at the bottom left corner.
}\label{fig:ratio_lhalir}
\end{figure}

In this section, we investigate if this evolution of the obscuration properties of the starbursts in U\slash LIRGs is already present in our intermediate-$z$ U\slash LIRGs. To do so, in Figure~\ref{fig:ratio_lhalir} we plot the ratio between the SFR derived from the observed H$\alpha$ emission and from the total IR luminosity. These SFRs are computed using equations 2 and 4 of \citet{Murphy2011}, respectively, which we reproduce below:

\begin{equation}
{\rm SFR_{H\alpha}\slash M_\odot\,yr^{-1}}=5.37\times10^{-42}\left(L_{\rm H\alpha}\slash {\rm erg\,s^{-1}}\right)
\end{equation}
\begin{equation}
{\rm SFR_{IR}\slash M_\odot\,yr^{-1}}=3.88\times10^{-44}\left(L_{\rm IR}\slash {\rm erg\,s^{-1}}\right)
\end{equation}

The SFR$_{\rm H\alpha}$\slash SFR$_{\rm IR}$ ratio is a proxy for the average obscuration of the SF, assuming a constant SFR during the last $\sim$100\,Myr, since H$\alpha$ is only sensitive to the less obscured SF while the IR luminosity traces the obscured SF (see e.g., \citealt{Kennicutt2012}). For comparison, we plot the same ratio for the sample of local U\slash LIRGs of \citet{RodriguezZaurin2011} scaled to the \citet{Murphy2011} SFR calibrations.

{The average $\log$ SFR$_{\rm H\alpha}$\slash SFR$_{\rm IR}$ ratio of our intermediate-$z$ U\slash LIRGs is $-$2.2$\pm$0.4 while this ratio is $-$1.9$\pm$0.4 for local U\slash LIRGs. Therefore, our intermediate-$z$ U\slash LIRGs have average dust extinctions slightly higher than those of local U\slash LIRGs for the same $L_{\rm IR}$ but within the scatter observed in local objects.}
Therefore, we do not find evidence for the expected decrease of the dust extinction in these intermediate-$z$ U\slash LIRGs. This suggests that the SF taking place in these intermediate-$z$ galaxies resembles the SF of local U\slash LIRGs in terms of obscuration. 
To fully confirm that the SF in local and intermediate-$z$ U\slash LIRGs is similar, we will need direct measurements of the size of the SF regions through high-angular resolution data.

\subsection{Metallicity}

In this section, we use the [\ion{N}{ii}]658nm\slash H$\alpha$ ratio (N2 index) to compare the metalicity of the SWIFT U\slash LIRGs and that of local U\slash LIRGs. The N2 index is well correlated with the metal abundance of galaxies (e.g., \citealt{Pettini2004, Marino2013}). However, this optical index may be subject to extinction effects, specially in dusty systems like U\slash LIRGs. \citet{Pereira2017Metal} used far-IR metallicity diagnostics, which are not affected by dust extinction, and found that, for local ULIRGs, optical and far-IR metallicity diagnostics provide compatible results.
Therefore, the results presented in this section should not be sensitive to the obscuration on these systems.

\begin{figure}
\centering
\includegraphics[width=0.4\textwidth]{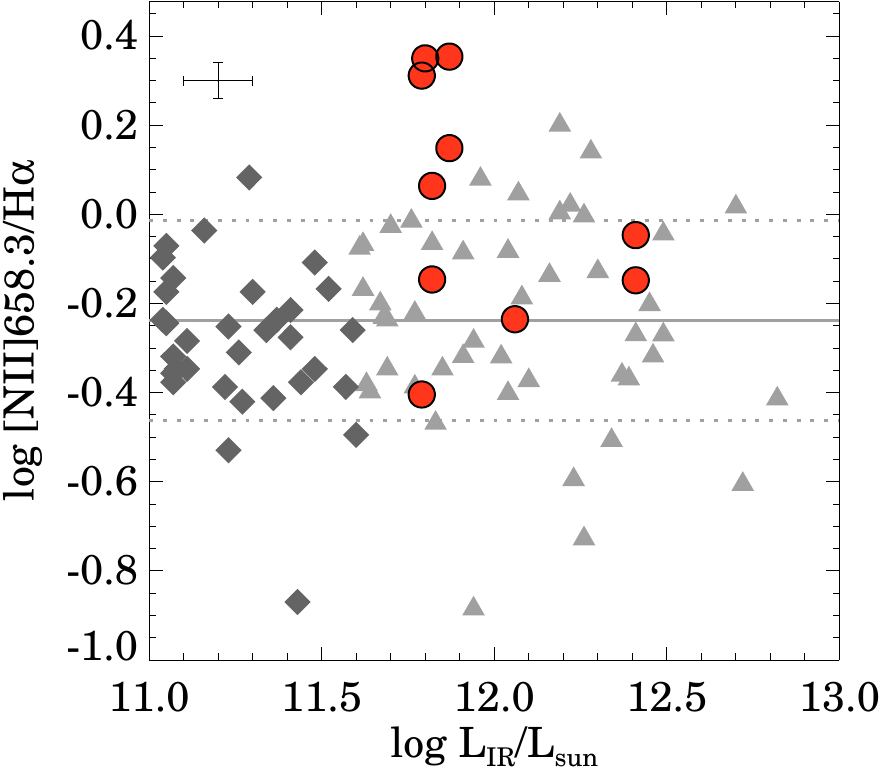}
\caption{\small [\ion{N}{ii}]658nm\slash H$\alpha$ ratio (N2 index) as a function the IR luminosity for the SWIFT U\slash LIRGs (red circles) and for local U\slash LIRGs {(gray triangles $\log L_{\rm IR}\slash L_\odot > 11.6$; and dark gray diamonds $\log L_{\rm IR}\slash L_\odot < 11.6$)}. The data for the local U\slash LIRGs are from \citet{Rupke2008}, \citet{AAH09PMAS}, and \citet{MonrealIbero2010}. The solid gray line marks the mean N2 index for local U\slash LIRGs and the dashed line its $\pm$1$\sigma$ confidence range. The 2 SWIFT U\slash LIRGs with broad H$\alpha$ profiles (SWIRE3 and BOOTES1) are not included in this figure. The average uncertainties are indicated by the error bars at the top left corner.
}\label{fig:ratio_n2}
\end{figure}

We plot the N2 index in Figure~\ref{fig:ratio_n2} as a function of the IR luminosity for the SWIFT U\slash LIRGs and for local U\slash LIRGs from \citet{Rupke2008}, \citet{AAH09PMAS}, and \citet{MonrealIbero2010}. SWIFT sources with broad emission lines (i.e., AGN) are excluded from this plot. We find that the SWIFT U\slash LIRGs have N2 indexes (1.1$\pm$0.8) slightly higher than local U\slash LIRGs (0.7$\pm$0.3).
However, this difference is not statistically significant. 

\subsection{Dust temperature}\label{ss:td}

The sample of intermediate-$z$ U\slash LIRGs studied in this {work} was selected based on their observed 250\micron\ flux. This wavelength is considerably longer than the 60\micron\ flux used for the selection of U\slash LIRGs both locally \citep{Kim1998b} and at intermediate-$z$ \citep{Combes2011} based on {\it IRAS} observations. Therefore, the 250\micron\ selection might be more sensitive to { selecting} galaxies with colder dust than samples selected at shorter wavelengths.

\begin{figure}
\centering
\includegraphics[width=0.4\textwidth]{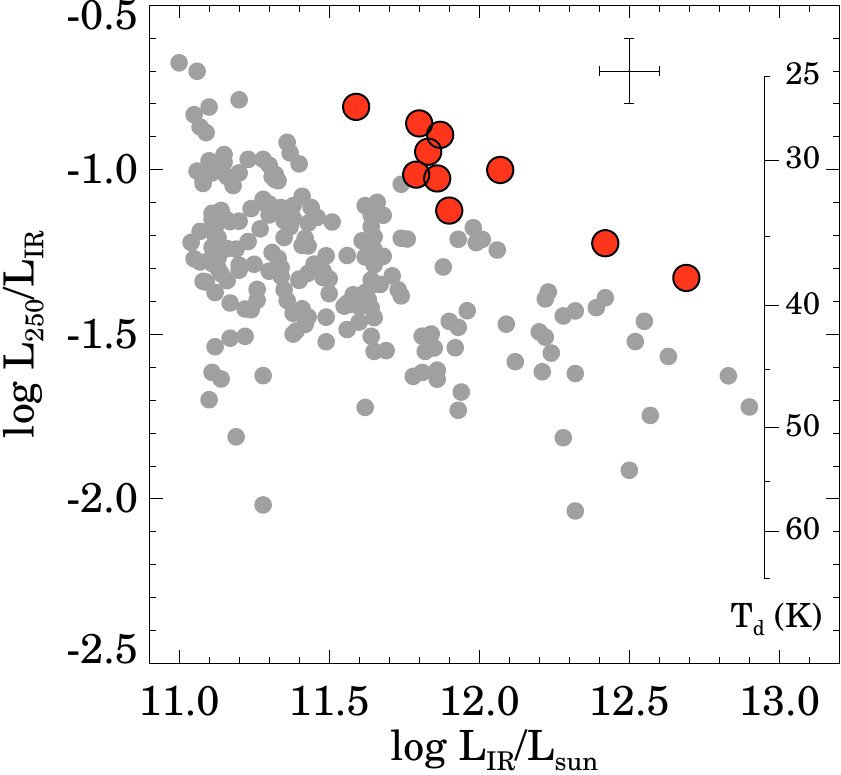}
\caption{\small $L_{\rm 250}$\slash $L_{\rm IR}$ ratio as a function of the total IR luminosity ($L_{\rm IR}$). The red circles represent our sample of intermediate-z U\slash LIRGs. The gray circles correspond to local U\slash LIRGs from \citet{Chu2017}, \citet{Sturm2011}, and \citet{Farrah2013}. The dust temperature ($T_{\rm d}$) scale is calculated assuming a gray body with $\beta=1.6$ (see Section~\ref{ss:td} for details). The average uncertainties are indicated by the error bars at the top right corner.}\label{fig:lir_250}
\end{figure}

To investigate this, we compare the $L_{\rm 250}$\slash $L_{\rm IR}$ ratio measured in our intermediate-$z$ U\slash LIRGs and that of local U\slash LIRGs (Figure~\ref{fig:lir_250}). For the intermediate-$z$ galaxies, we use the \textit{Herschel}\slash SPIRE 250\micron\ fluxes from \citet{Magdis2014}. For the local U\slash LIRGs, we take SPIRE 250\micron\ fluxes, distances, and $L_{\rm IR}$ published by \citet{Chu2017}. This includes most of the local U\slash LIRGs in the Revised Bright Galaxy Sample \citep{SandersRBGS}. In addition, we obtained the SPIRE 250\micron\ fluxes for another 16 ULIRGs observed by \textit{Herschel} \citep{Sturm2011, Farrah2013} from the SPIRE Point Source Catalog \citep{Schulz2017}. A k correction was applied to the observed 250\micron\ fluxes for all the galaxies assuming a dust temperature of 40\,K. This correction varies $<$15\% for dust temperatures between 25 and 50\,K and reduces the 250\micron\ fluxes by 15\% and 50\% on average for the local and the intermediate-$z$ ULIRGs respectively.

The $L_{\rm 250}$\slash $L_{\rm IR}$ ratio is sensitive to the dust temperature. Assuming that the IR emission is well represented by a gray-body with $\beta$=1.6 \citep{Casey2012}, we calculated the $L_{\rm 250}$\slash $L_{\rm IR}$ ratio as a function of the dust temperature (see Figure~\ref{fig:lir_250}). We find that most of the U\slash LIRGs have ratios compatible with dust temperatures between 25 and 50\,K. Also, we observe the known trend of increasing dust temperatures with increasing $L_{\rm IR}$ (e.g., \citealt{Hwang2010}).

{We find that these intermediate-$z$ U\slash LIRGs have $L_{\rm 250}$\slash $L_{\rm IR}$ ratios higher than local U\slash LIRGs with the same $L_{\rm IR}$. This suggests that the average dust temperature in these systems is lower than in local U\slash LIRGs. Therefore, the selection of our sample at 250\micron\ reveals a population of cold intermediate-$z$ \hbox{U\slash LIRGs} different from local U\slash LIRGs and warm \hbox{intermediate-$z$} \hbox{U\slash LIRGs} selected at 60\micron\ \citep{Combes2011}. 
}

\section{Conclusions}\label{s:conclusions}

We compared the properties of a sample of 9 intermediate-$z$ ($0.2<z<0.4$) U\slash LIRG systems with those of local U\slash LIRGs using optical IFS data. We aim to determine if the differences between local U\slash LIRGs and $z>1$ U\slash LIRGs are already present at $0.2<z<0.4$.
We used the H$\alpha$+[\ion{N}{ii}] emission to model the 2D kinematics and to measure the N2 metallicity index. These new results for the intermediate-$z$ U\slash LIRGs, combined with their IR properties ($L_{\rm IR}$ and $L_{\rm 250\mu m}$), were used to compare the dynamical state, SFR obscuration, metallicity, and dust temperature of local and intermediate-$z$ U\slash LIRGs. The main results of this paper are the following:

\begin{enumerate}
\item The optical IFS data reveal that 4 out 9 U\slash LIRGs are composed by two individual objects located at projected distances between 20 and 30\,kpc. Therefore, they are at an early stage of the interaction. The remaining 5 objects in our sample appear as isolated galaxies. However, higher angular resolution imaging is needed to identify, if present, signs of recent interaction\slash merger events in these 5 objects.
In two systems, we find broad H$\alpha$ profiles $\sigma=300-400$\,km\,s$^{-1}$ (FWHM$\sim750-900$\,km\,s$^{-1}$) in their integrated spectra suggesting the presence of an AGN.

\item The corrected for redshift effects velocity dispersion ($\sigma_{\rm z_0}$) and dynamical ratio ($v\slash \sigma_{\rm z_0}$) of these intermediate-$z$ U\slash LIRGs indicate that half of them (4 out of 8) belong to interacting systems and 1 or 2 out of 8 are compatible with being isolated galaxies. For most cases (6 out of 8), this classification is in agreement with that obtained from the images.

\item The ratio between the obscured SFR traced by the IR luminosity and the un-obscured SFR traced by H$\alpha$ of the intermediate-$z$ U\slash LIRGs is compatible with the range measured in local U\slash LIRGs with the same $L_{\rm IR}$. Therefore, the obscuration properties of the SF regions in intermediate-$z$ U\slash LIRGs seem to be comparable to those of local U\slash LIRGs.

\item The average N2 metallicity index ([\ion{N}{ii}]658nm\slash H$\alpha$ ratio) of the intermediate-$z$ U\slash LIRGs is slightly higher (1.1$\pm$0.8) than that of local U\slash LIRGs (0.7$\pm$0.3). Although this difference is not statistically significant.

\item {We find that the ratio between the 250\micron\ luminosity and the total $L_{\rm IR}$ of these intermediate-$z$ U\slash LIRGs is higher than that of local U\slash LIRGs with the same $L_{\rm IR}$. This is compatible with reduced dust temperatures in these intermediate-$z$ U\slash LIRGs compared to local U\slash LIRGs. This, together with the enhanced molecular gas content traced by their CO emission (see \citealt{Magdis2014}), suggest different ISM conditions in these intermediate-$z$ and local U\slash LIRGs.}

\end{enumerate}

In summary, the main differences between these intermediate-$z$ U\slash LIRGs and local U\slash LIRGs are: (1) intermediate-$z$ U\slash LIRGs seem to be at an earlier interaction stage; and (2) the dust temperature is lower in these U\slash LIRGs selected at 250\micron. On the contrary, the obscuration level of the SFR seems to be similar in both local and these intermediate-$z$ U\slash LIRGs.

\section*{Acknowledgements}
{We thank the anonymous referee for useful comments and suggestions. }
MPS and DR acknowledge support from STFC through grant ST/N000919/1. MPS, NT, FC, MT and MR acknowledge support from STFC through grant ST/N002717/1.

\begin{landscape}
\begin{small}
\begin{table}
\centering
\caption{H$\alpha$ and [\ion{N}{ii}]658.3\,nm emission properties in galaxies with a broad component.}
\label{tab:fluxes_broad}
\begin{tabular}{@{}lcccccccccccc@{}}
\hline
Name & $\lambda_{\rm H\alpha}$\,$^{a}$ & F(H$\alpha_{\rm narrow}$)$^{b}$ & $\frac{\rm H\alpha_{\rm broad}}{\rm H\alpha_{\rm narrow}}$\,$^{c}$ & $\frac{\rm [NII]658.3\,nm}{\rm H\alpha_{\rm narrow}}$\,$^{d}$ & $\frac{\rm [NII]658.3\,nm}{\rm H\alpha_{\rm broad}}$\,$^{e}$ & v$_{\rm narrow}$\,$^{f}$ & v$_{\rm broad}$\,$^{g}$ & $\sigma_{\rm narrow}$\,$^{h}$ & $\sigma_{\rm broad}$\,$^{i}$ & $L$(H$\alpha_{\rm narrow}$)\,$^{j}$  \\
& (nm) & (10$^{-16}$\,erg\,cm$^{-2}$\,s$^{-1}$) & & & & (km\,s$^{-1}$) & (km\,s$^{-1}$) & (km\,s$^{-1}$) & (km\,s$^{-1}$) & (10$^{40}$\,erg\,s$^{-1}$) \\
\hline
SWIRE3 & 752.7 $\pm$ 0.8 & 11.50 $\pm$ 0.53 & 12.49 $\pm$ 0.58 & 0.87 $\pm$ 0.06 & 0.77 $\pm$ 0.01 & 40835 $\pm$ 44 & 41039 $\pm$ 51 & 71 $\pm$ 53 & 313 $\pm$ 80 & 6.7 \\
BOOTES1 & 886.8 $\pm$ 0.3 & 7.40 $\pm$ 0.13 & 1.77 $\pm$ 0.05 & 0.90 $\pm$ 0.02 & 2.80 $\pm$ 0.07 & 87614 $\pm$ 26 & 87507 $\pm$ 64 & 96 $\pm$ 34 & 377 $\pm$ 72 & 30.8 \\
\hline
\end{tabular}

\medskip
\raggedright \textbf{Notes:} 
$^{(a)}$ Observed wavelength of the narrow H$\alpha$ transition. 
$^{(b)}$ Observed flux of the narrow H$\alpha$ component. 
$^{(c)}$ Ratio between the broad and narrow H$\alpha$ fluxes. 
$^{(d,e)}$ Observed [NII]658.3\,nm\slash H$\alpha$ ratio for the narrow and broad components of these transitions, respectively.
$^{(f,g)}$  Velocity derived from the narrow and broad H$\alpha$ and [NII]658.3\,nm emission using the relativistic velocity definition, respectively.
$^{(h,i)}$ Width of the best-fit Gaussian profiles for the narrow and broad components, respectively.
$^{(j)}$ Observed luminosity of the H$\alpha$ narrow component.
\end{table}
\end{small}
\end{landscape}

\clearpage

\appendix

\section{SWIFT close pairs emission maps}\label{apx:emission_pairs}

\begin{figure*}[!h]
\centering
\includegraphics[width=0.9\textwidth]{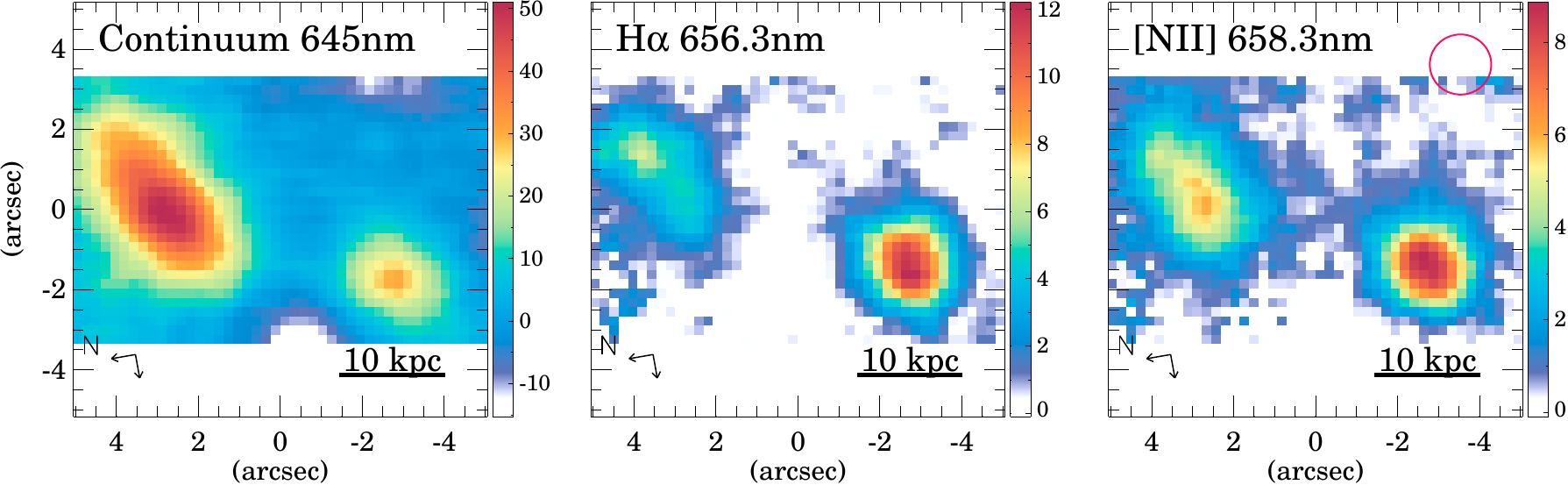}
\caption{\small Same as Figure~\ref{fig:system_map_xmm2} but for CDFS2.}\label{fig:system_map_cdfs2}
\end{figure*}

\begin{figure*}
\centering
\includegraphics[width=0.9\textwidth]{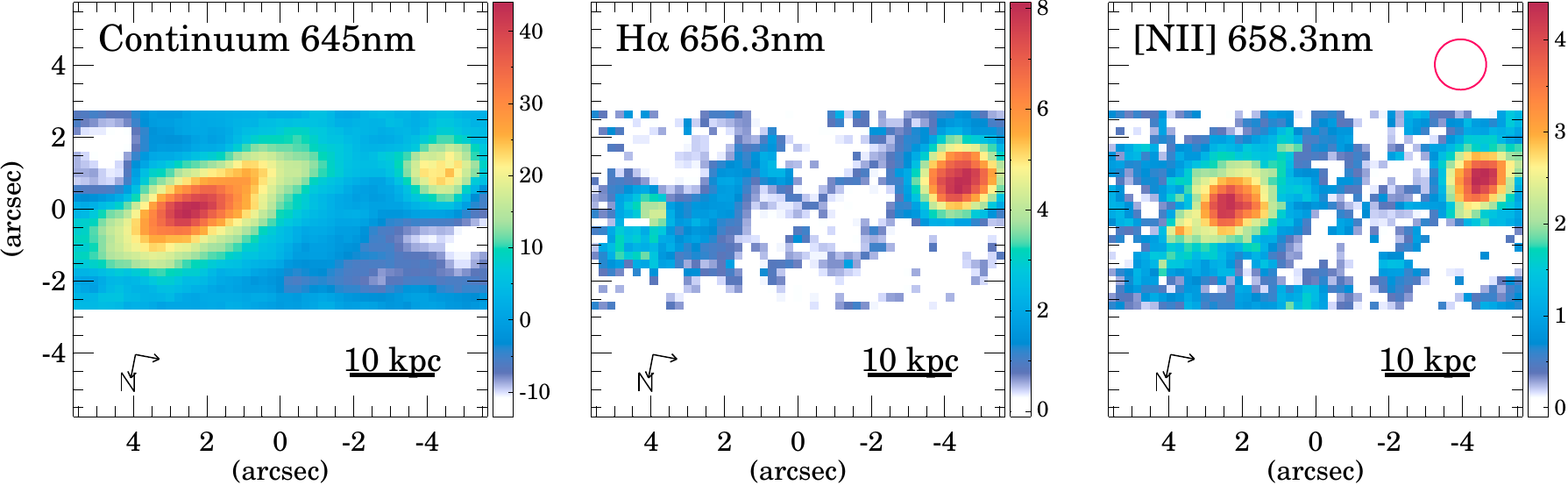}
\caption{\small Same as Figure~\ref{fig:system_map_xmm2} but for CDFS1.}
\end{figure*}

\begin{figure*}
\centering
\includegraphics[width=0.9\textwidth]{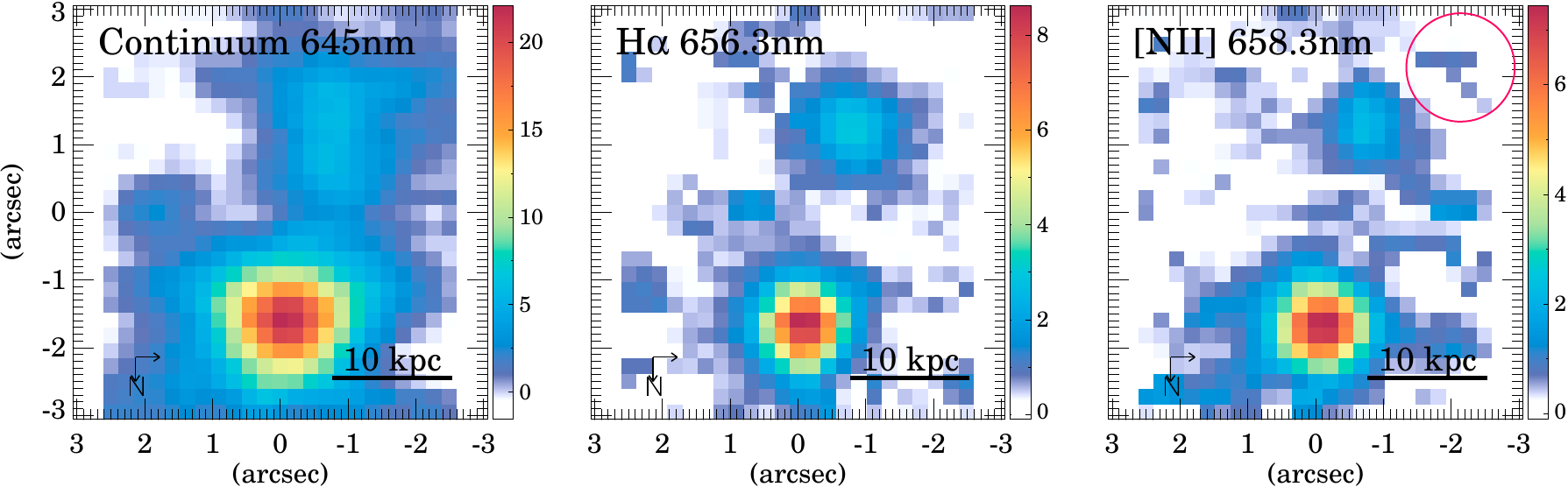}
\caption{\small Same as Figure~\ref{fig:system_map_xmm2} but for FLS02.}\label{fig:system_map_fls02}
\end{figure*}

\clearpage
\section{SWIFT individual emission maps}\label{apx:emission_indiv}

\begin{figure*}
\centering
\includegraphics[width=0.8\textwidth]{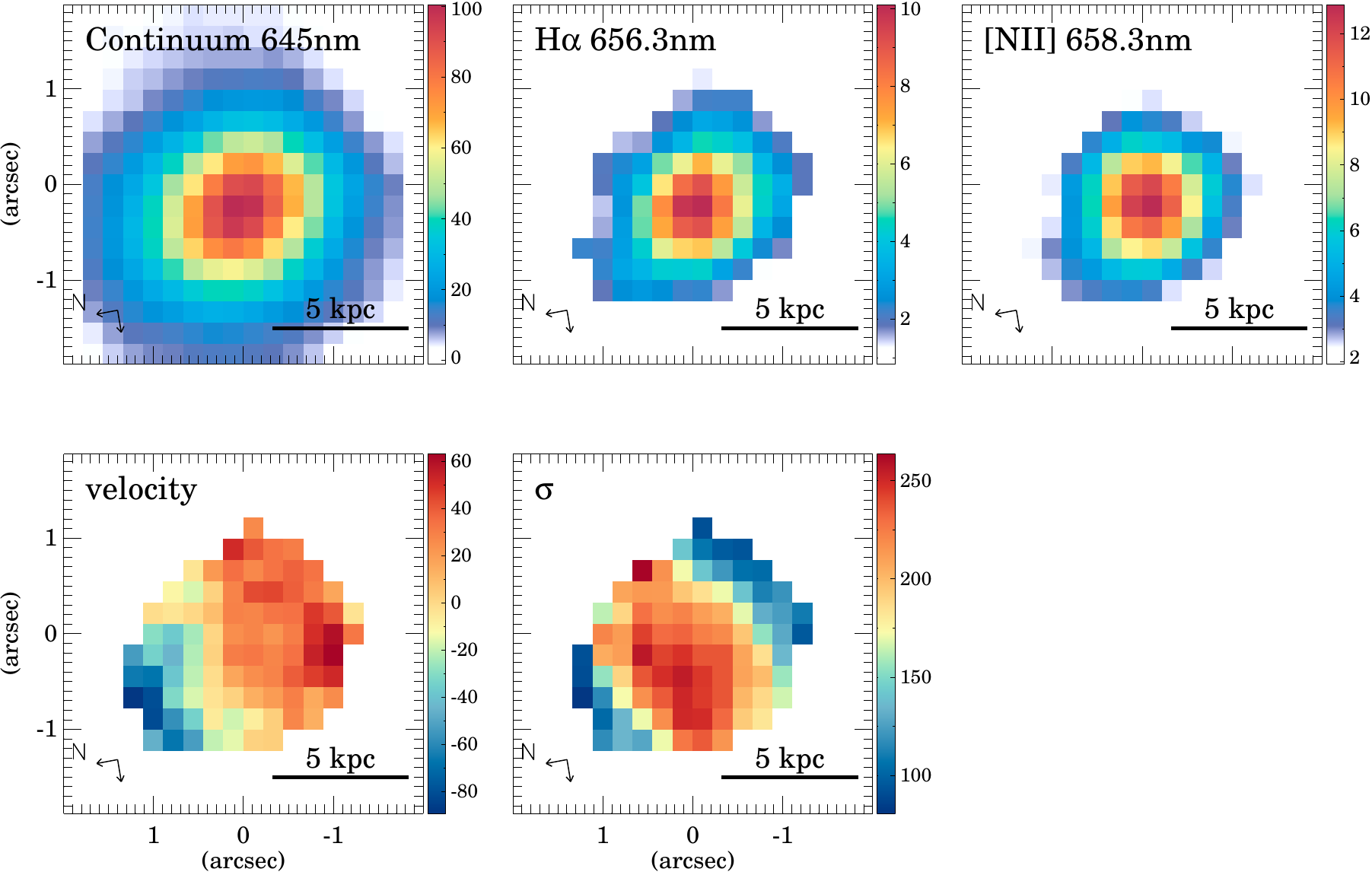}
\caption{\small Same as Figure~\ref{fig:map_xmm2w} but for XMM2 E.}
\end{figure*}

\begin{figure*}
\centering
\includegraphics[width=0.8\textwidth]{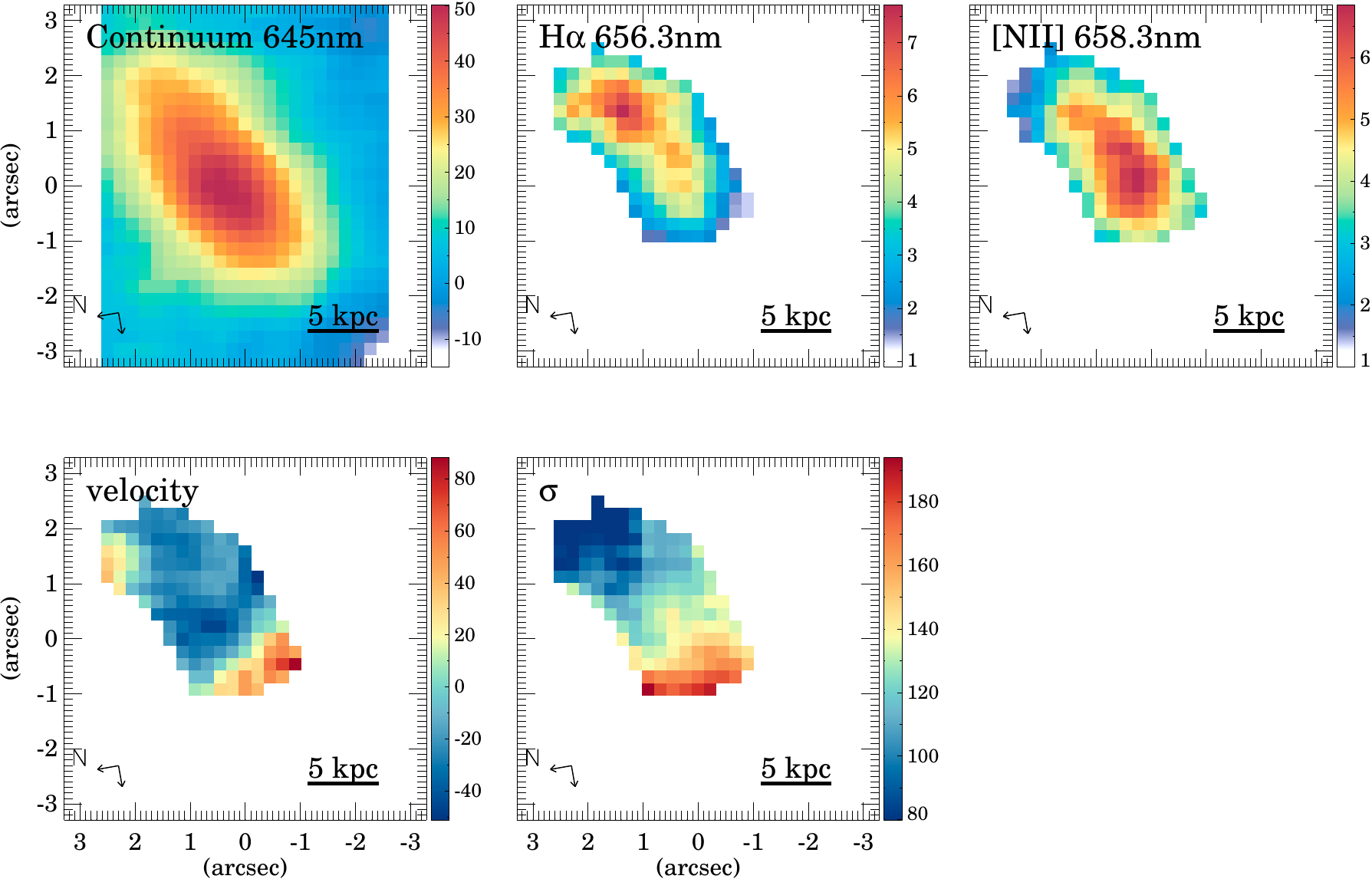}
\caption{\small Same as Figure~\ref{fig:map_xmm2w} but for CDFS2 N.}\label{fig:map_cdfs2n}
\end{figure*}

\begin{figure*}
\centering
\includegraphics[width=0.8\textwidth]{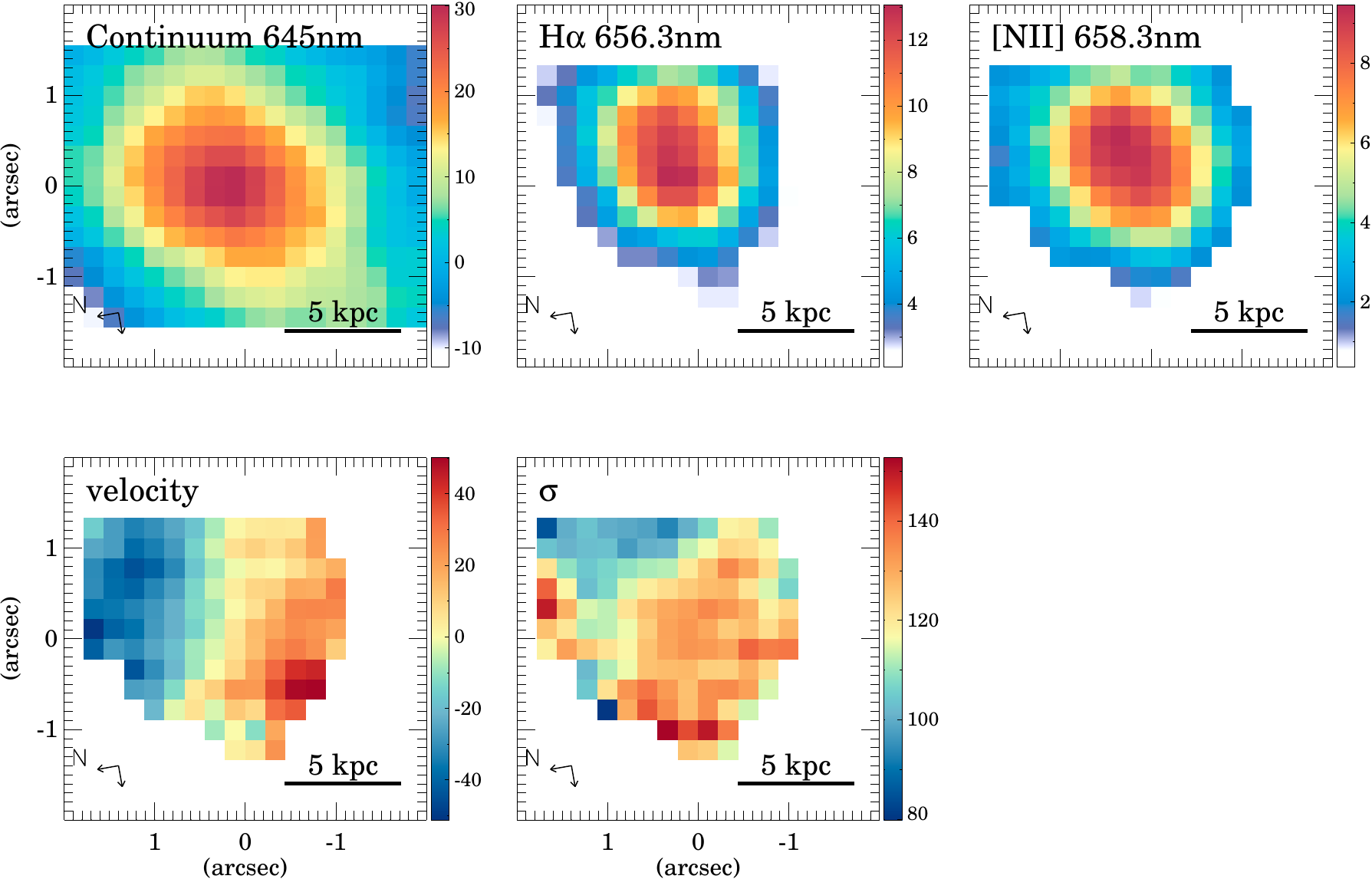}
\caption{\small Same as Figure~\ref{fig:map_xmm2w} but for CDFS2 S.}
\end{figure*}

\begin{figure*}
\centering
\includegraphics[width=0.8\textwidth]{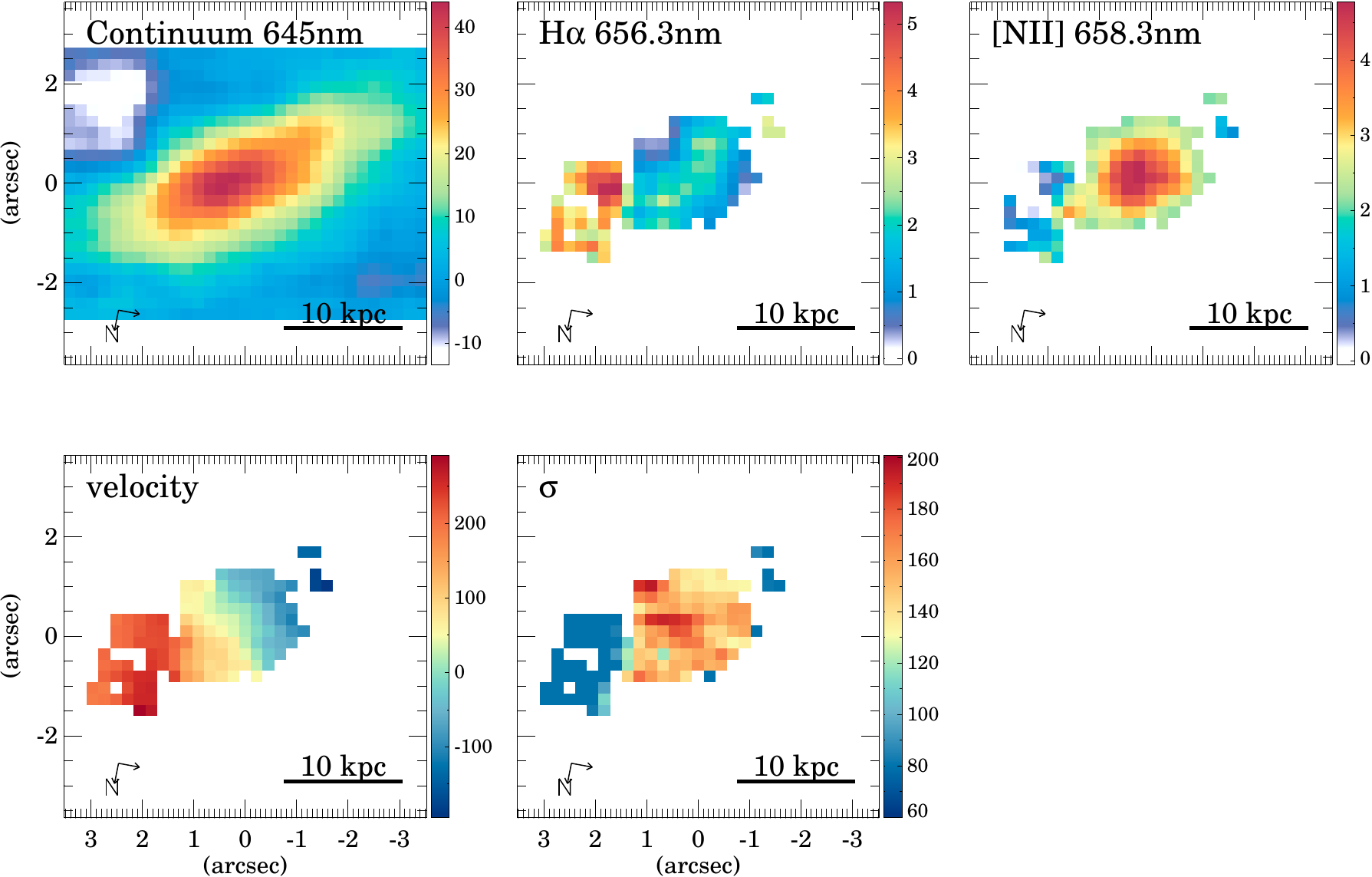}
\caption{\small Same as Figure~\ref{fig:map_xmm2w} but for CDFS1 W.}
\end{figure*}

\begin{figure*}
\centering
\includegraphics[width=0.8\textwidth]{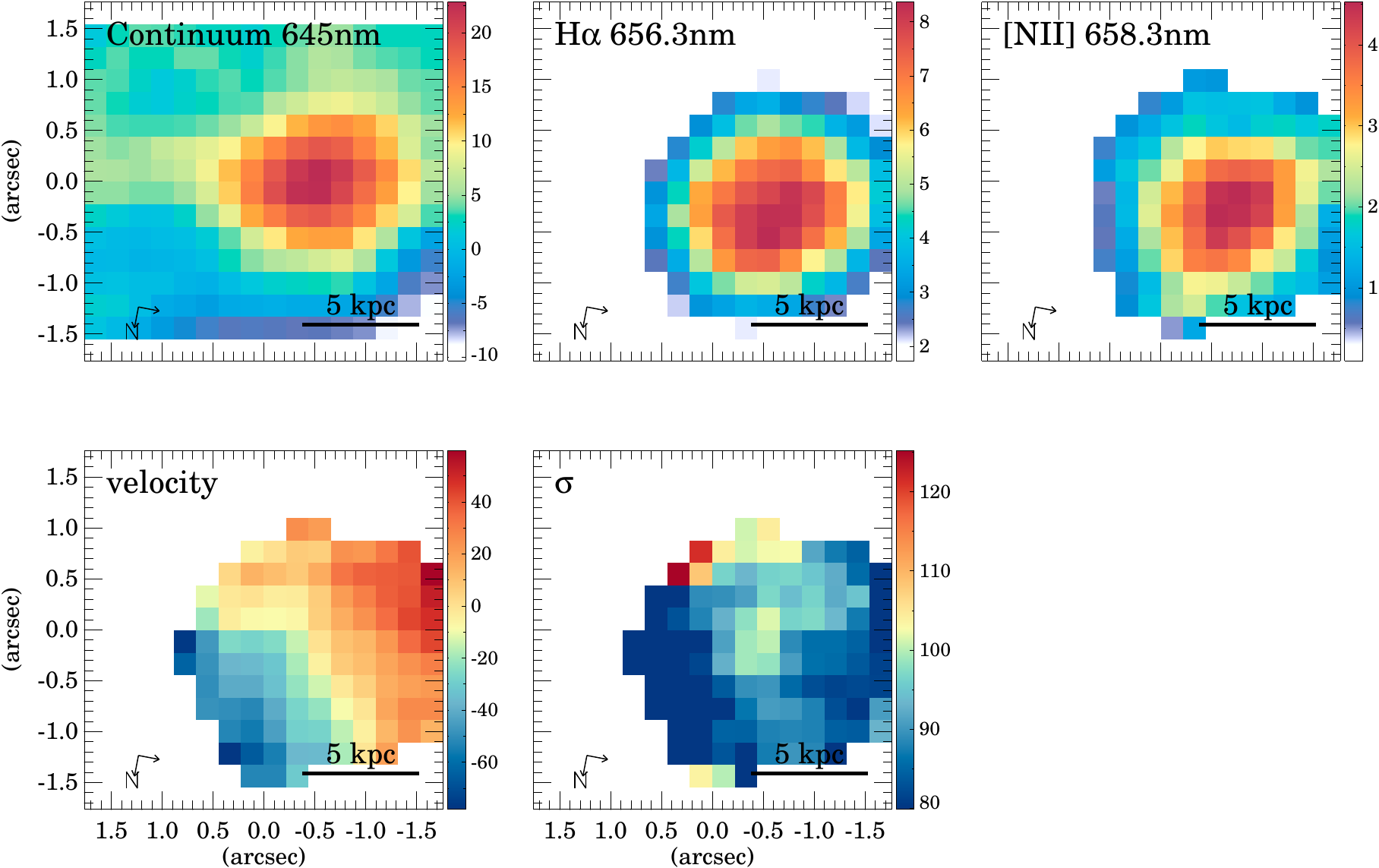}
\caption{\small Same as Figure~\ref{fig:map_xmm2w} but for CDFS1 E.}
\end{figure*}

\begin{figure*}
\centering
\includegraphics[width=0.8\textwidth]{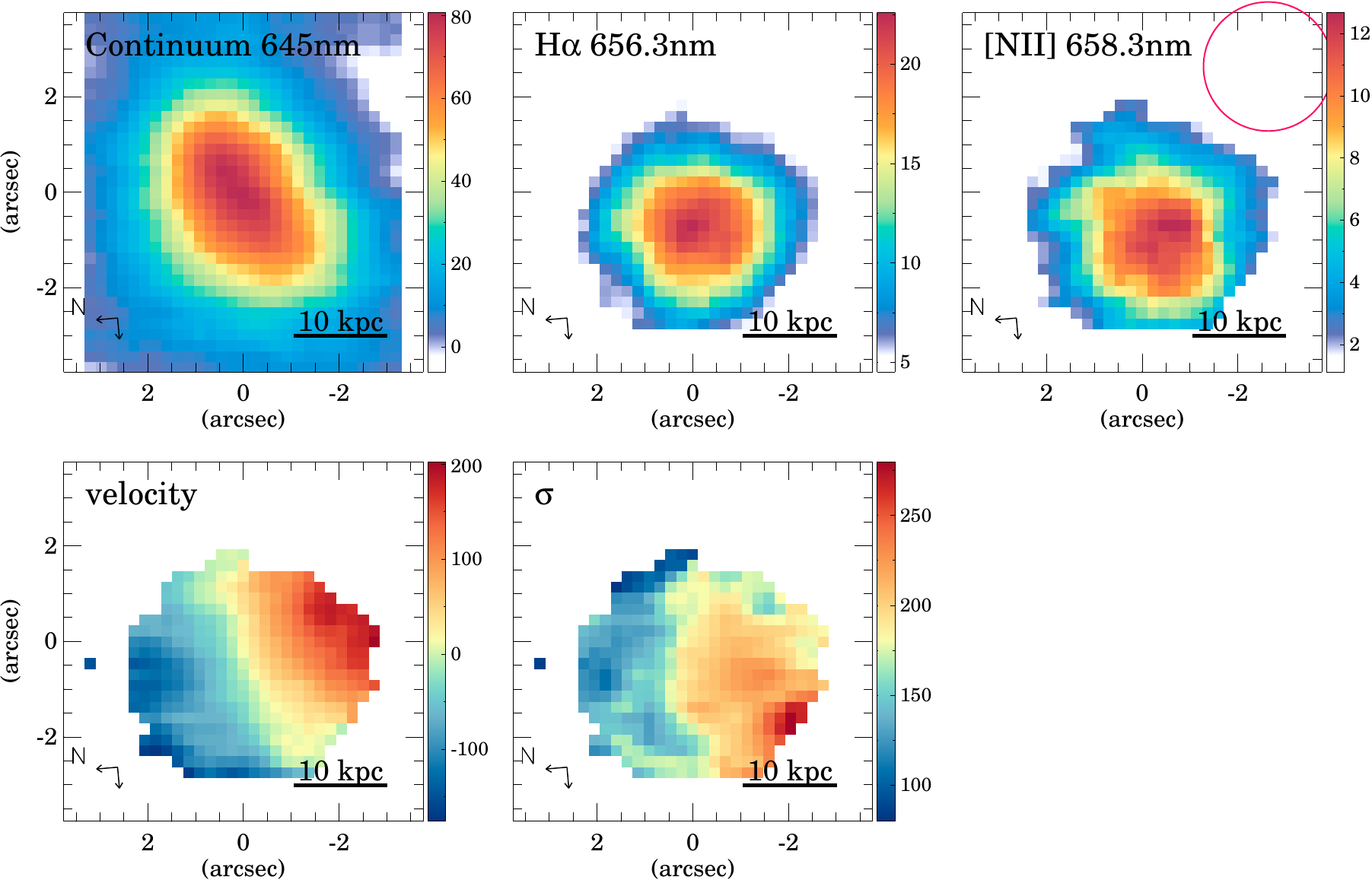}
\caption{\small Same as Figure~\ref{fig:map_xmm2w} but for SWIRE5. Note that the SWIRE5 W H$\alpha$ and [\ion{N}{ii}] emission are not included in the maps because of the lower redshift of this component (0.195) compared to the redshift (0.366) of the main IR emitter of the system (SWIRE5 E). The red circle in the right-hand panel corresponds to the seeing FWHM.}
\end{figure*}

\begin{figure*}
\centering
\includegraphics[width=0.8\textwidth]{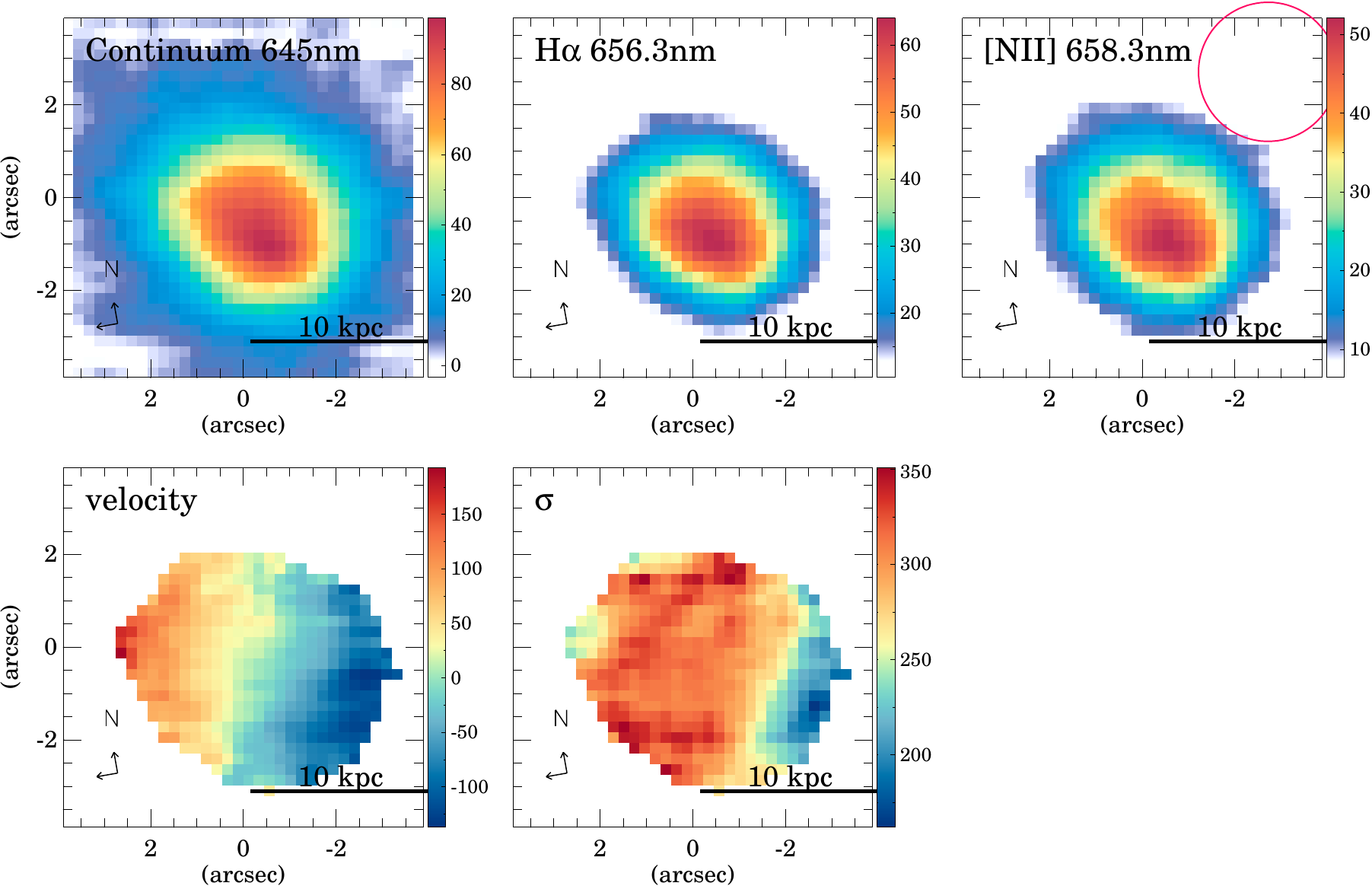}
\caption{\small Same as Figure~\ref{fig:map_xmm2w} but for SWIRE3. The red circle in the right-hand panel corresponds to the seeing FWHM.}
\end{figure*}

\begin{figure*}
\centering
\includegraphics[width=0.8\textwidth]{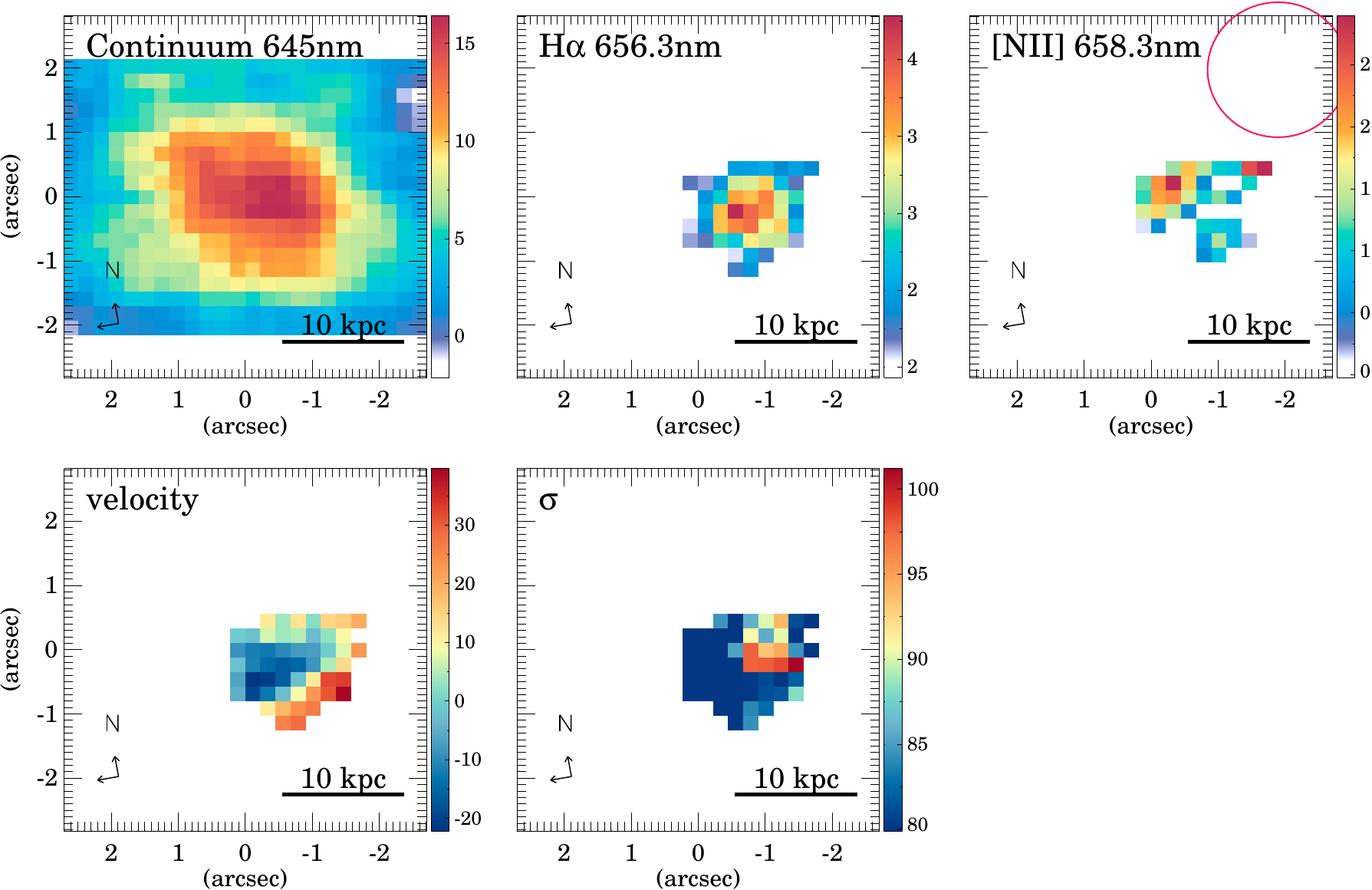}
\caption{\small Same as Figure~\ref{fig:map_xmm2w} but for SWIRE7. The red circle in the right-hand panel corresponds to the seeing FWHM.}
\end{figure*}

\begin{figure*}
\centering
\includegraphics[width=0.8\textwidth]{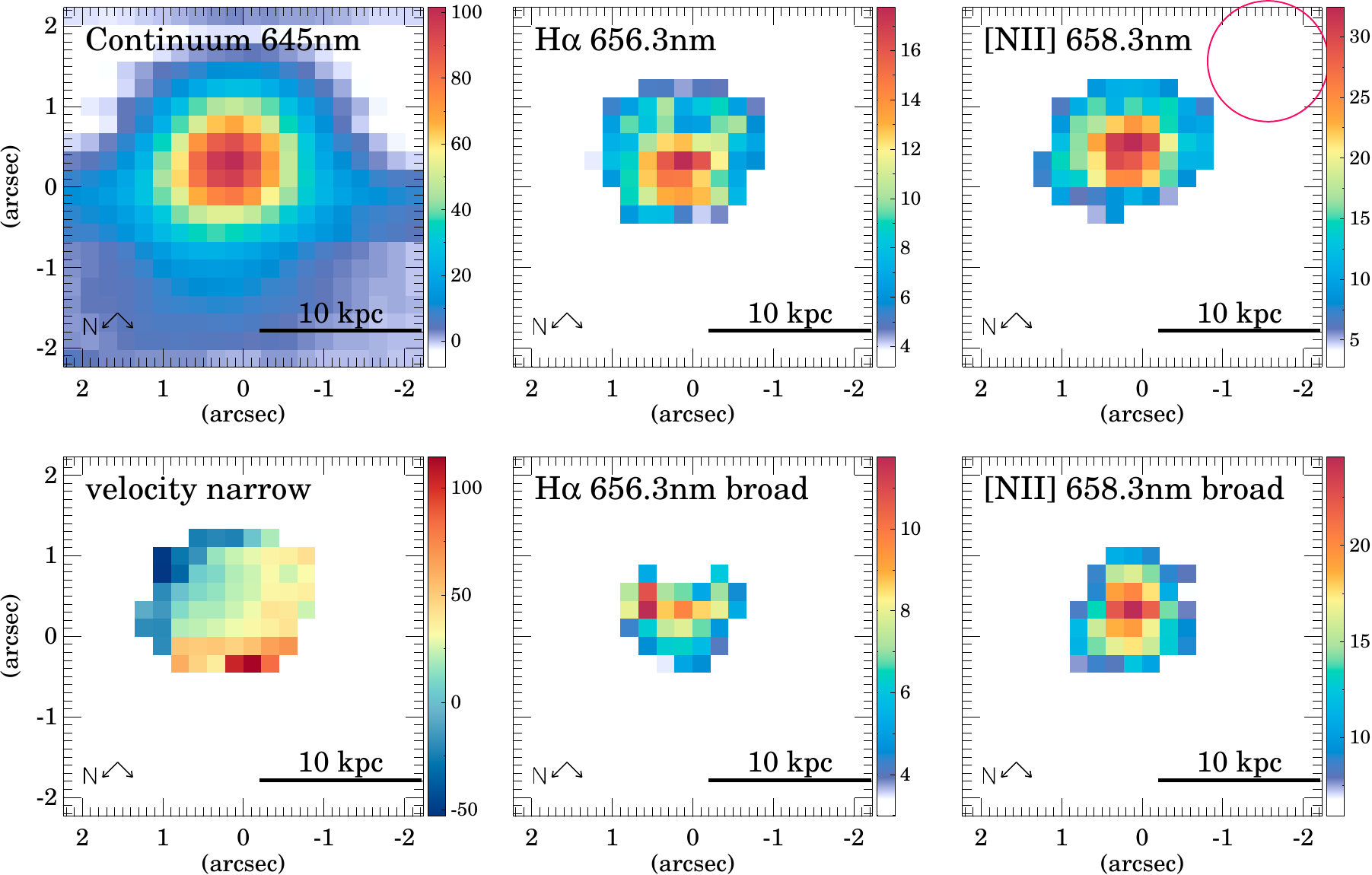}
\caption{\small Same as Figure~\ref{fig:map_xmm2w} but for BOOTES1. The velocity dispersion is not well constrained so this map is not included here. The lower panels also show the distribution of the broad H$\alpha$ (middle), and [\ion{N}{ii}] (right) emissions. The red circle in the right-hand panel corresponds to the seeing FWHM.}
\end{figure*}

\begin{figure*}
\centering
\includegraphics[width=0.8\textwidth]{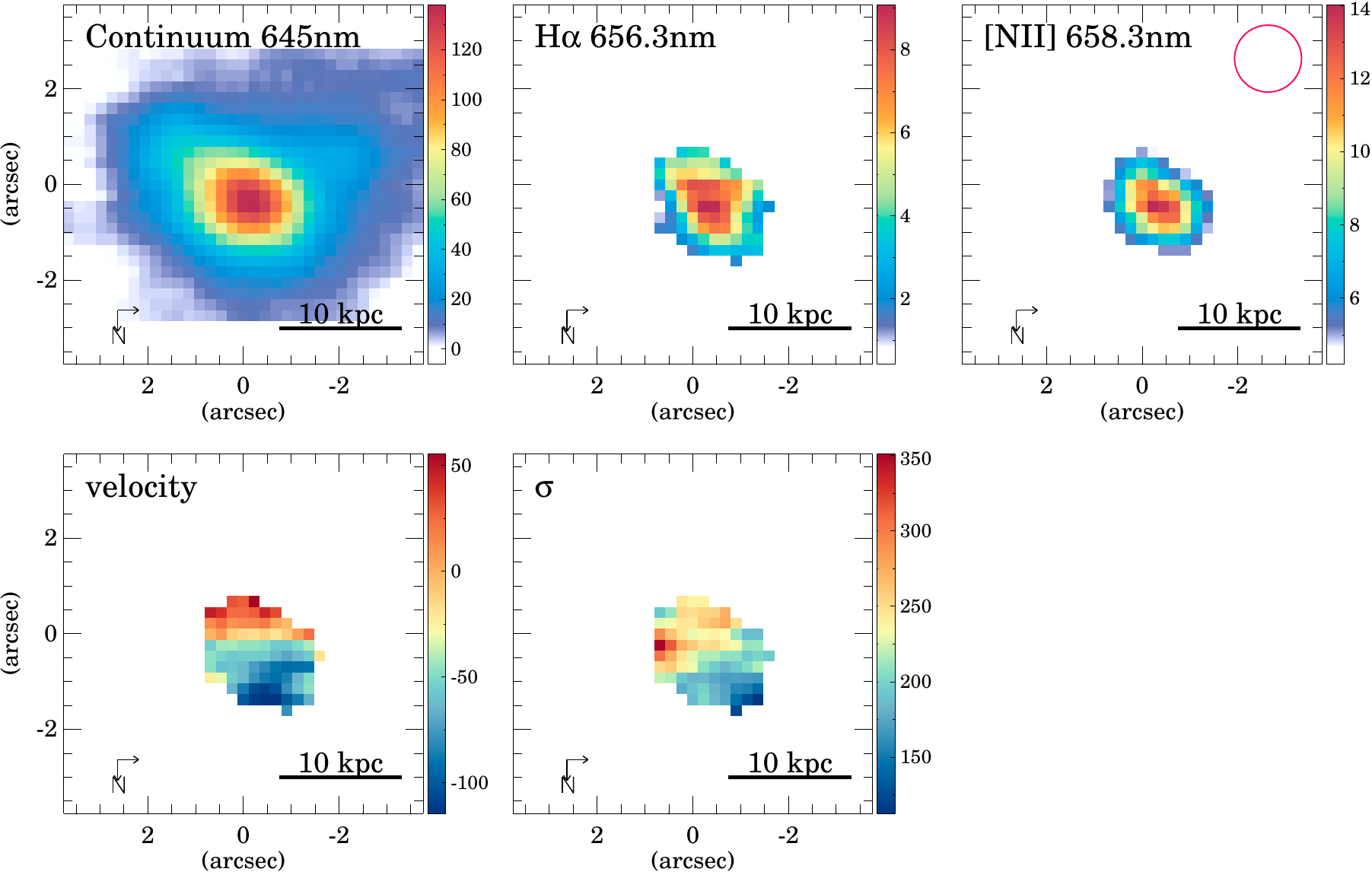}
\caption{\small Same as Figure~\ref{fig:map_xmm2w} but for BOOTES2. The red circle in the right-hand panel corresponds to the seeing FWHM.}
\end{figure*}

\begin{figure*}
\centering
\includegraphics[width=0.8\textwidth]{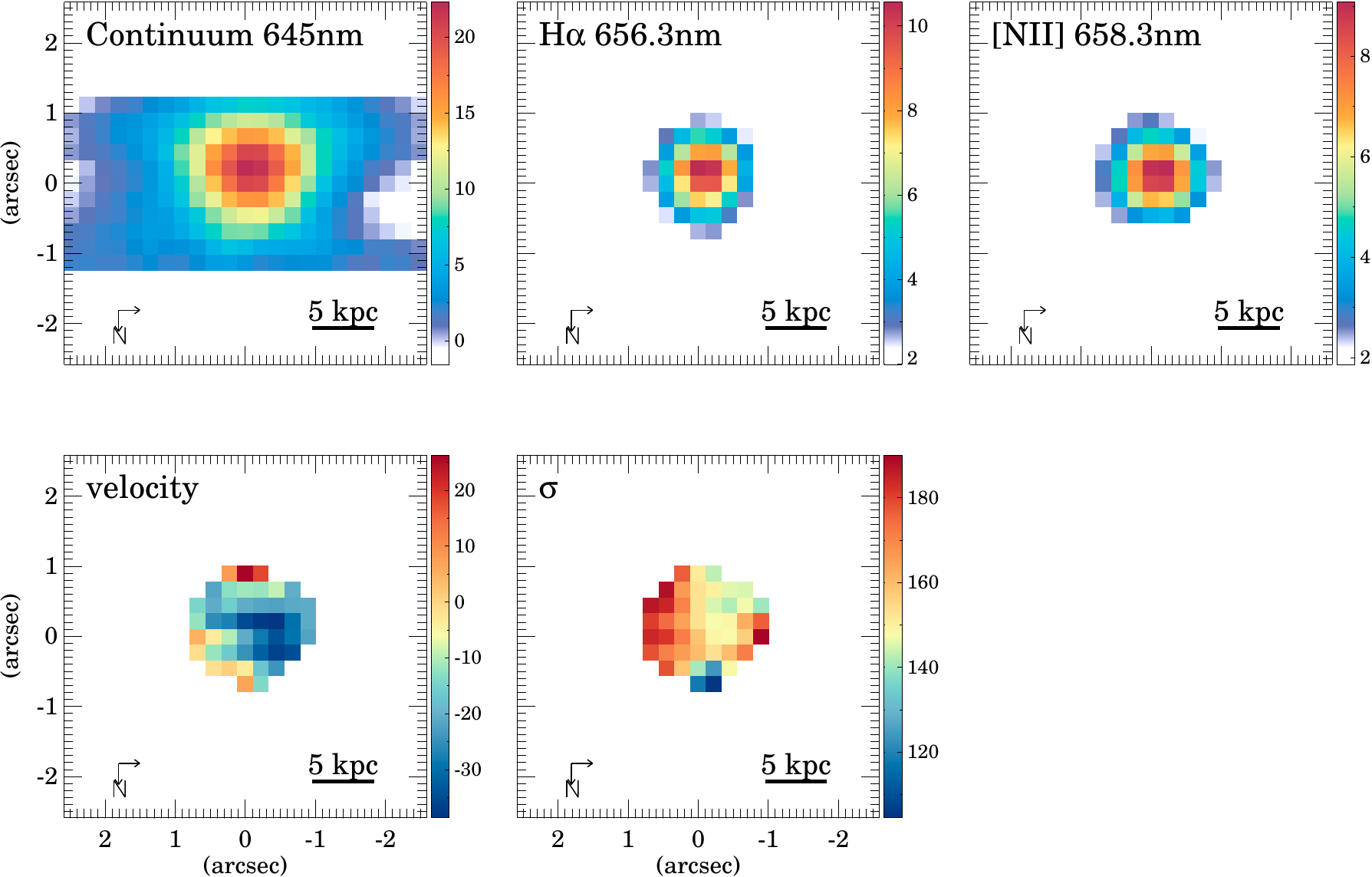}
\caption{\small Same as Figure~\ref{fig:map_xmm2w} but for FLS02 N.}\label{fig:map_fls02n}
\end{figure*}

\begin{figure*}
\centering
\includegraphics[width=0.8\textwidth]{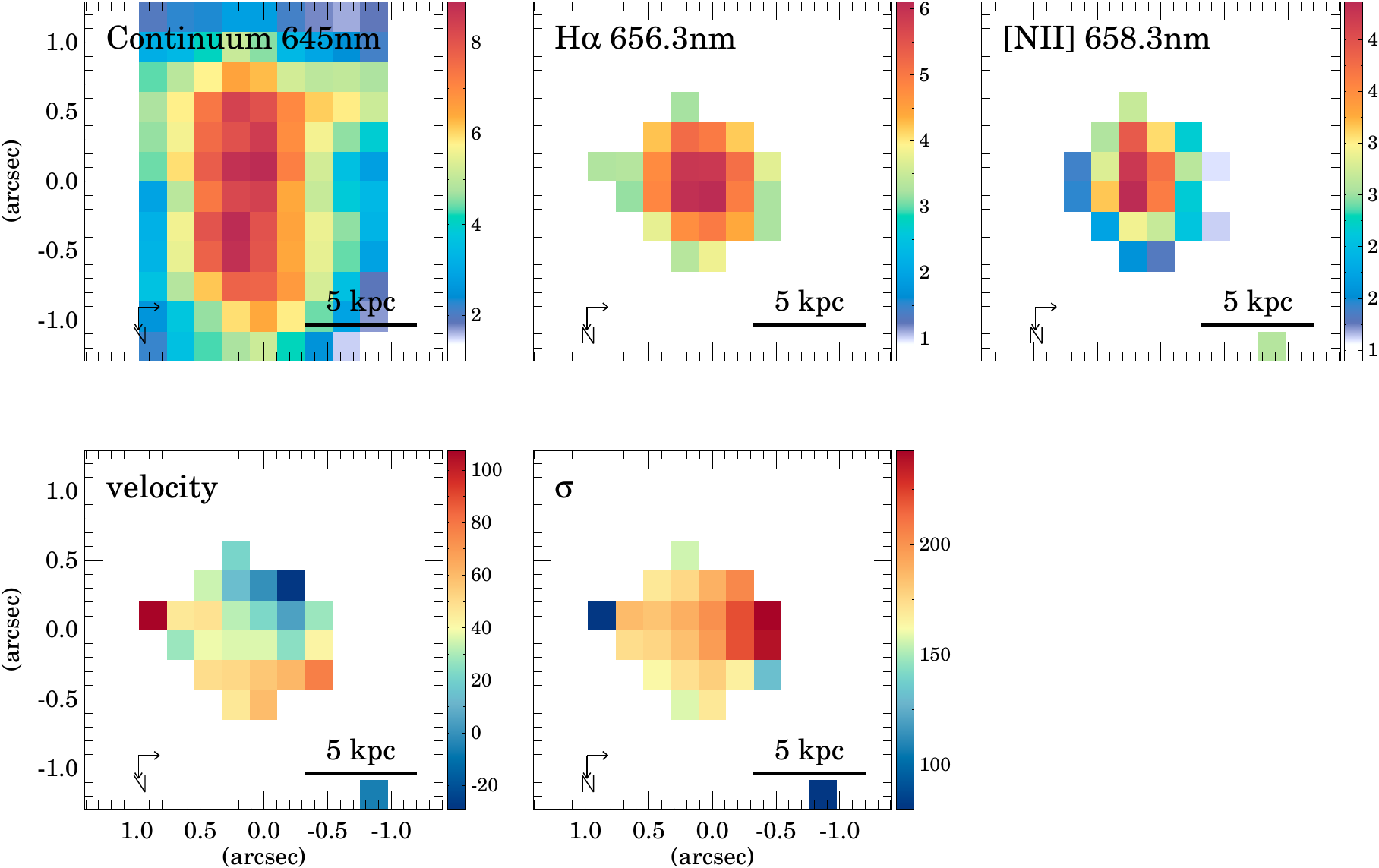}
\caption{\small Same as Figure~\ref{fig:map_xmm2w} but for FLS02 S.}\label{fig:map_fls02s}
\end{figure*}

\clearpage

\section{GalPak$^{\rm 3D}$ models}\label{apx:models}

\begin{figure*}
\centering
\includegraphics[width=0.68\textwidth]{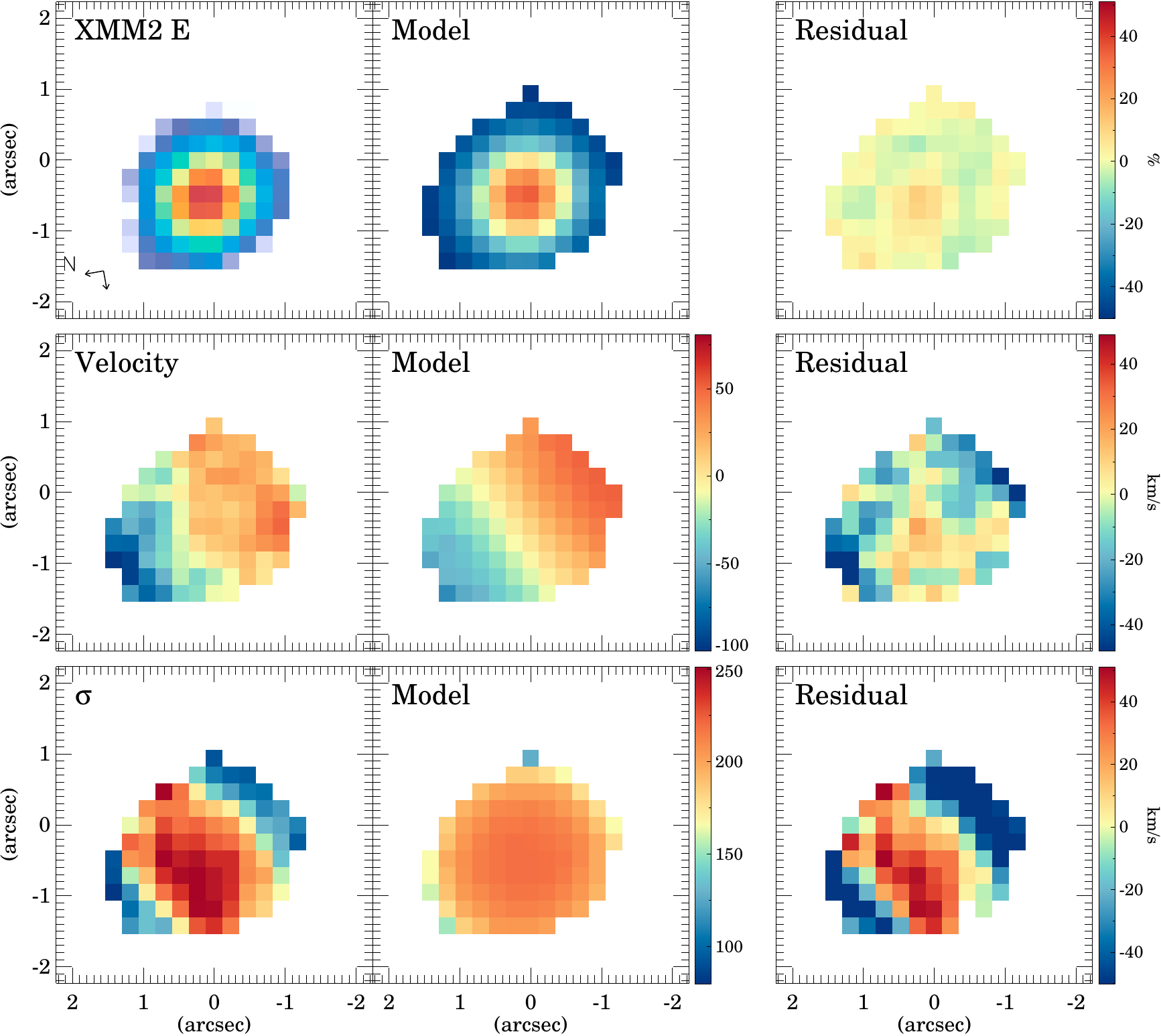}
\caption{\small Same as Figure~\ref{fig:galpak_xmm2w} but for XMM2 E.}
\end{figure*}

\begin{figure*}
\centering
\includegraphics[width=0.68\textwidth]{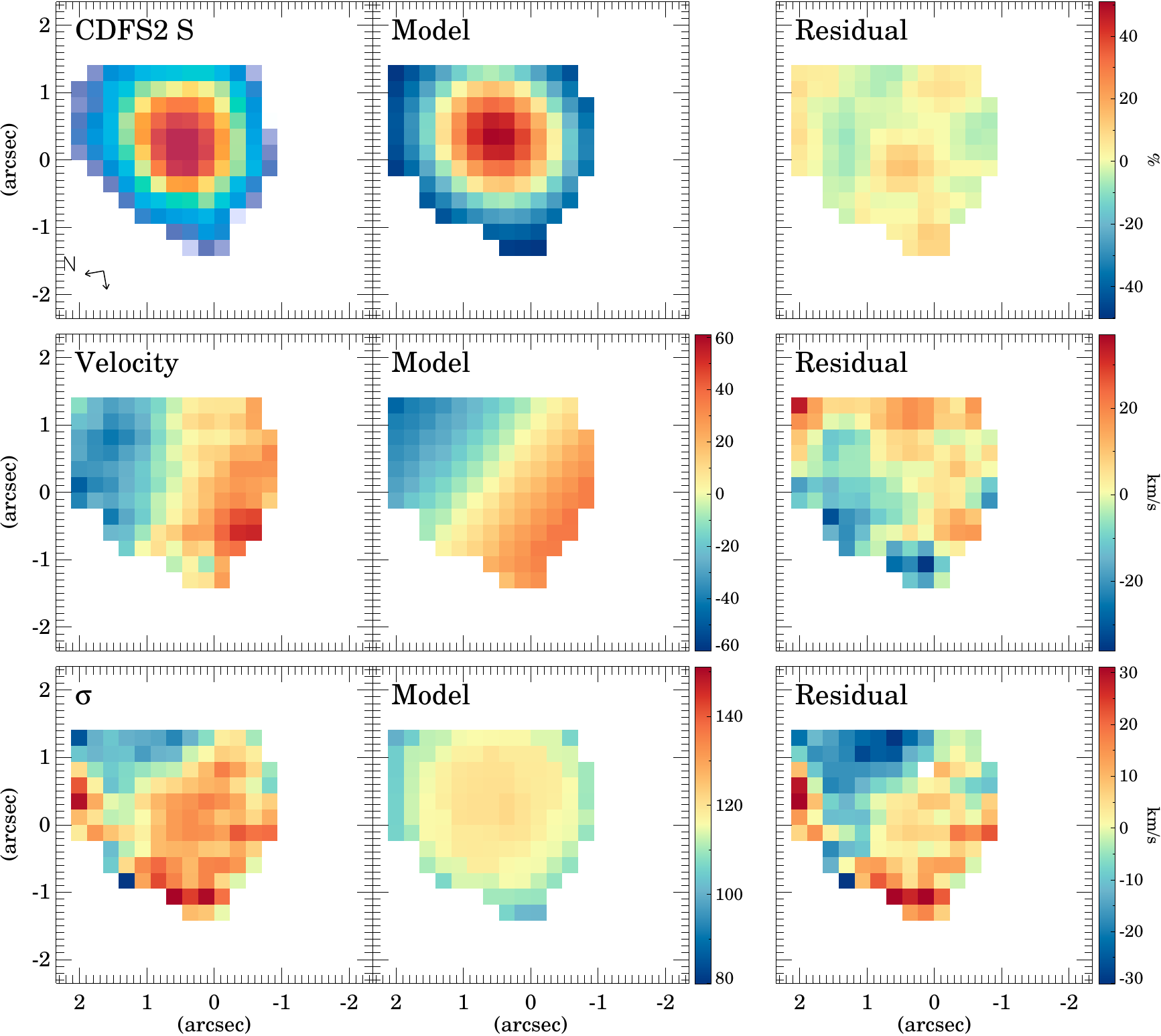}
\caption{\small Same as Figure~\ref{fig:galpak_xmm2w} but for CDFS2 S.}
\end{figure*}

\begin{figure*}
\centering
\includegraphics[width=0.68\textwidth]{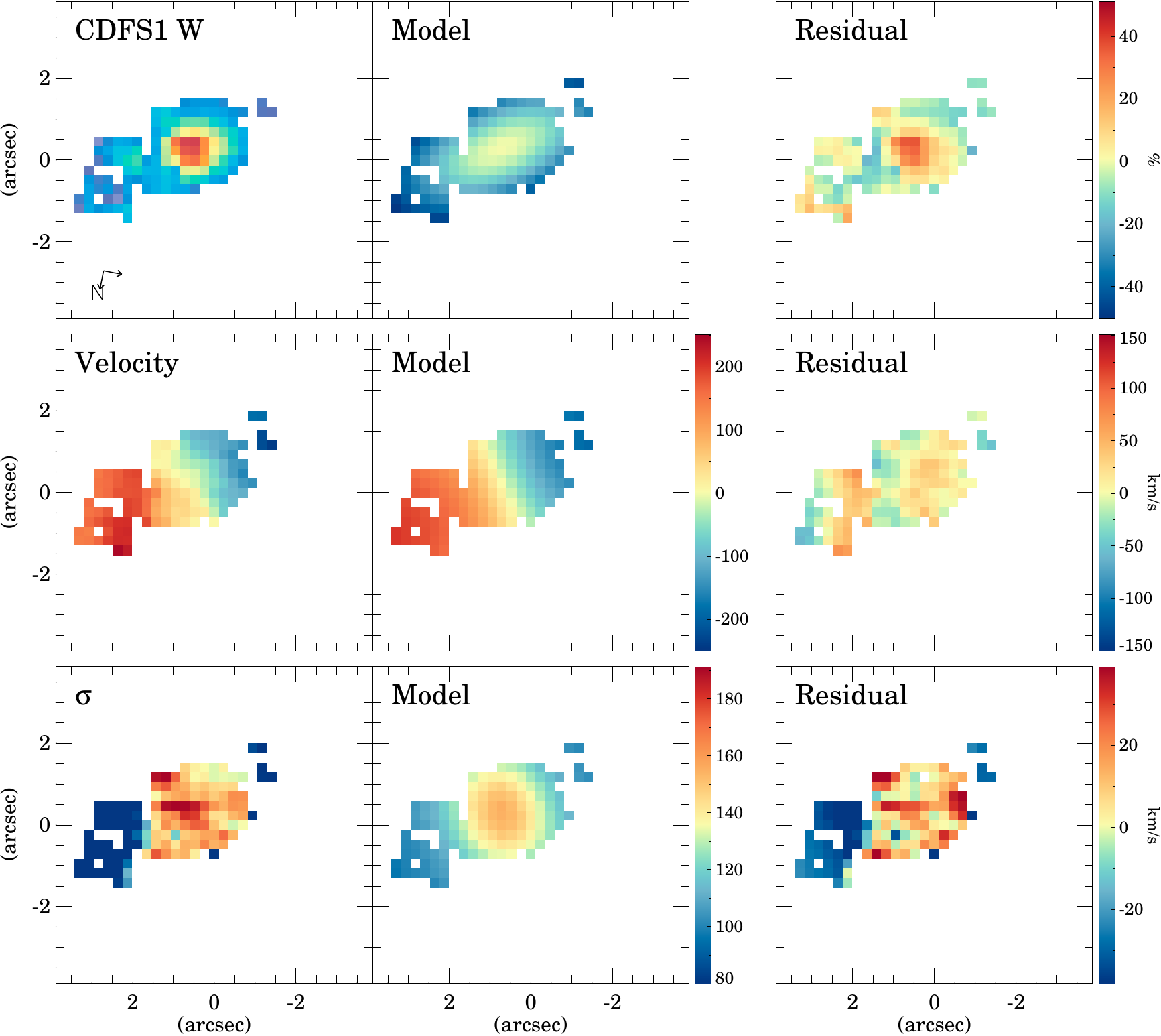}
\caption{\small Same as Figure~\ref{fig:galpak_xmm2w} but for CDFS1 W.}
\end{figure*}

\begin{figure*}
\centering
\includegraphics[width=0.68\textwidth]{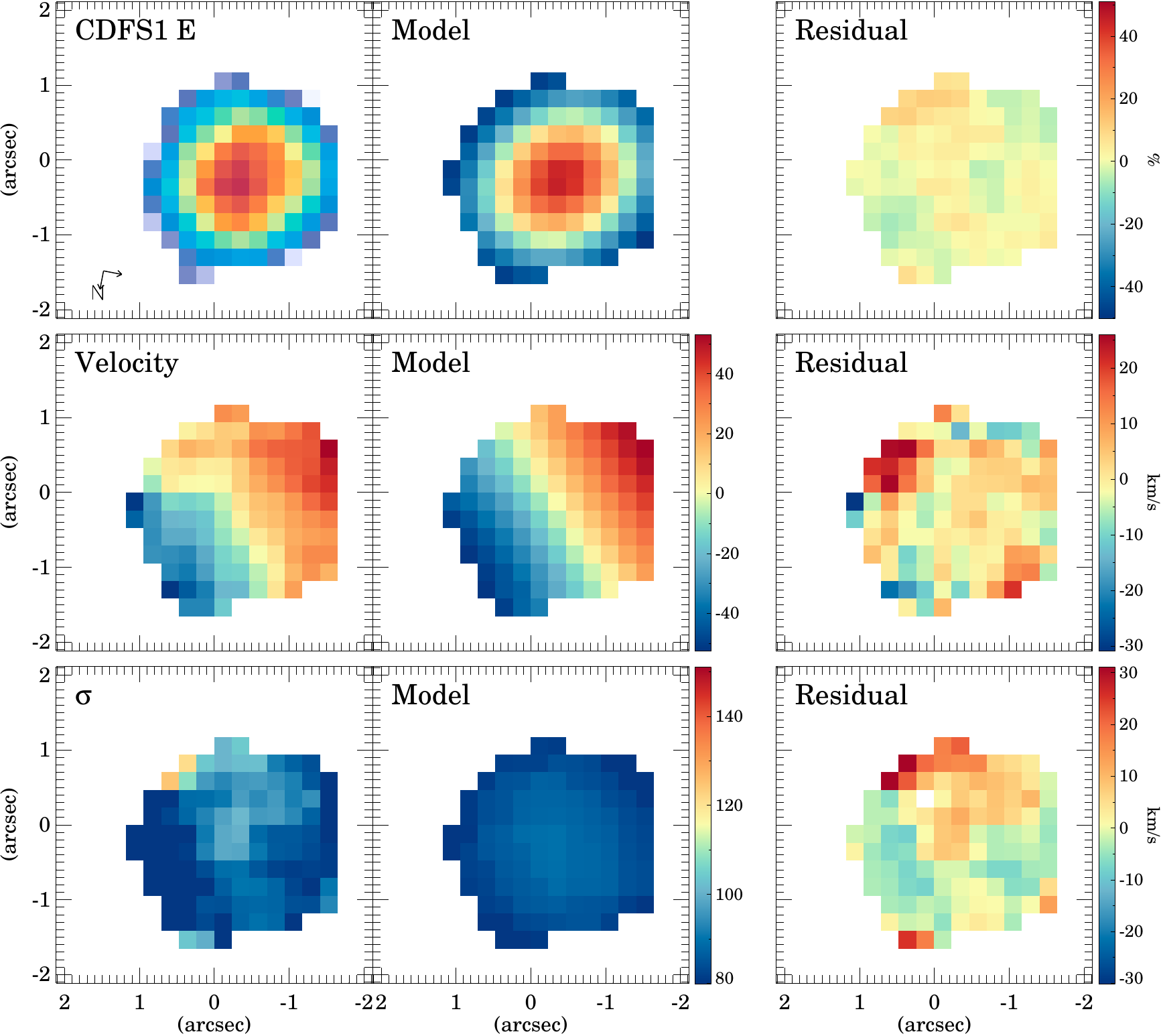}
\caption{\small Same as Figure~\ref{fig:galpak_xmm2w} but for CDFS1 E.}
\end{figure*}

\clearpage

\begin{figure*}
\centering
\includegraphics[width=0.68\textwidth]{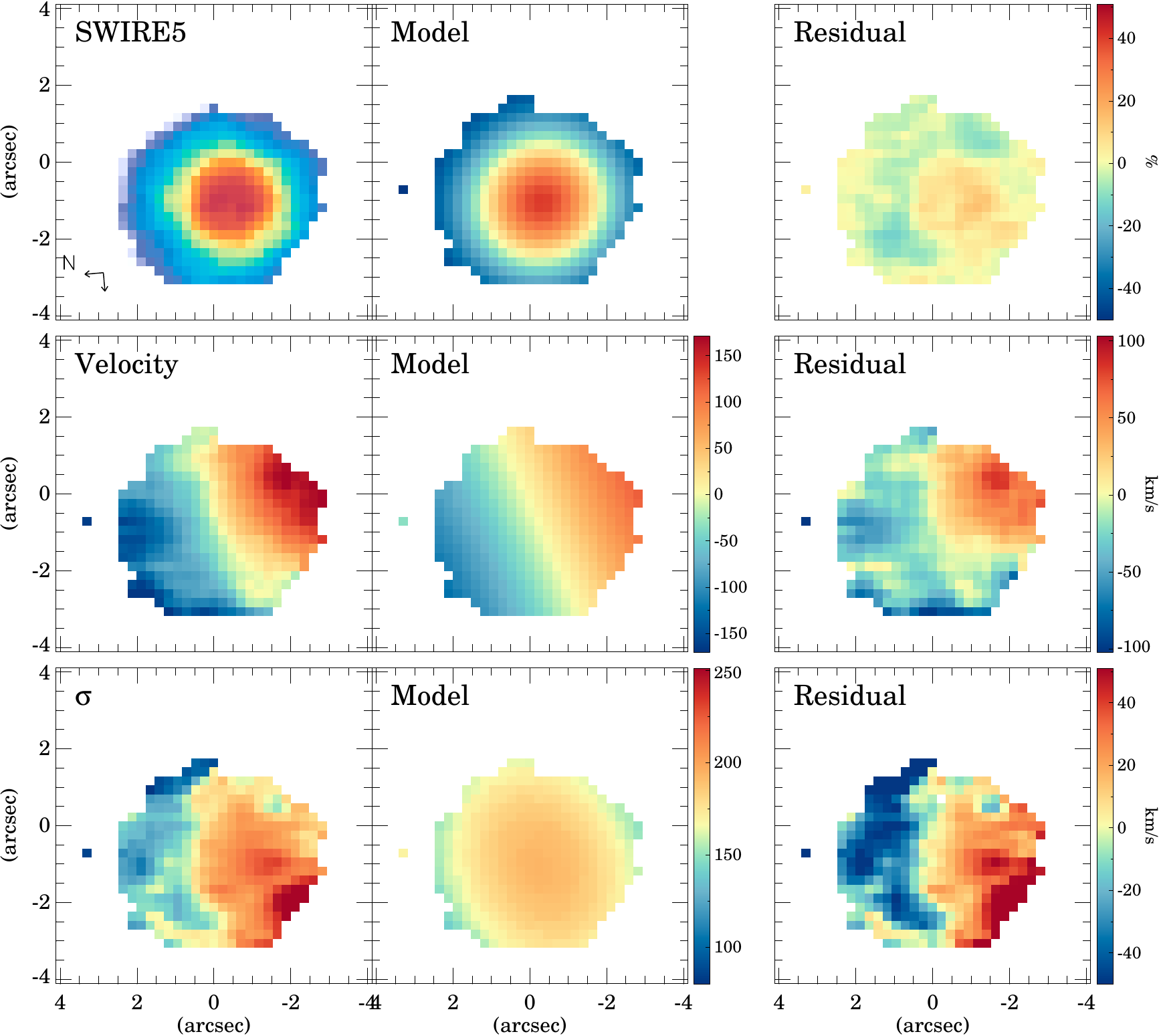}
\caption{\small Same as Figure~\ref{fig:galpak_xmm2w} but for SWIRE5.}
\end{figure*}

\begin{figure*}
\centering
\includegraphics[width=0.68\textwidth]{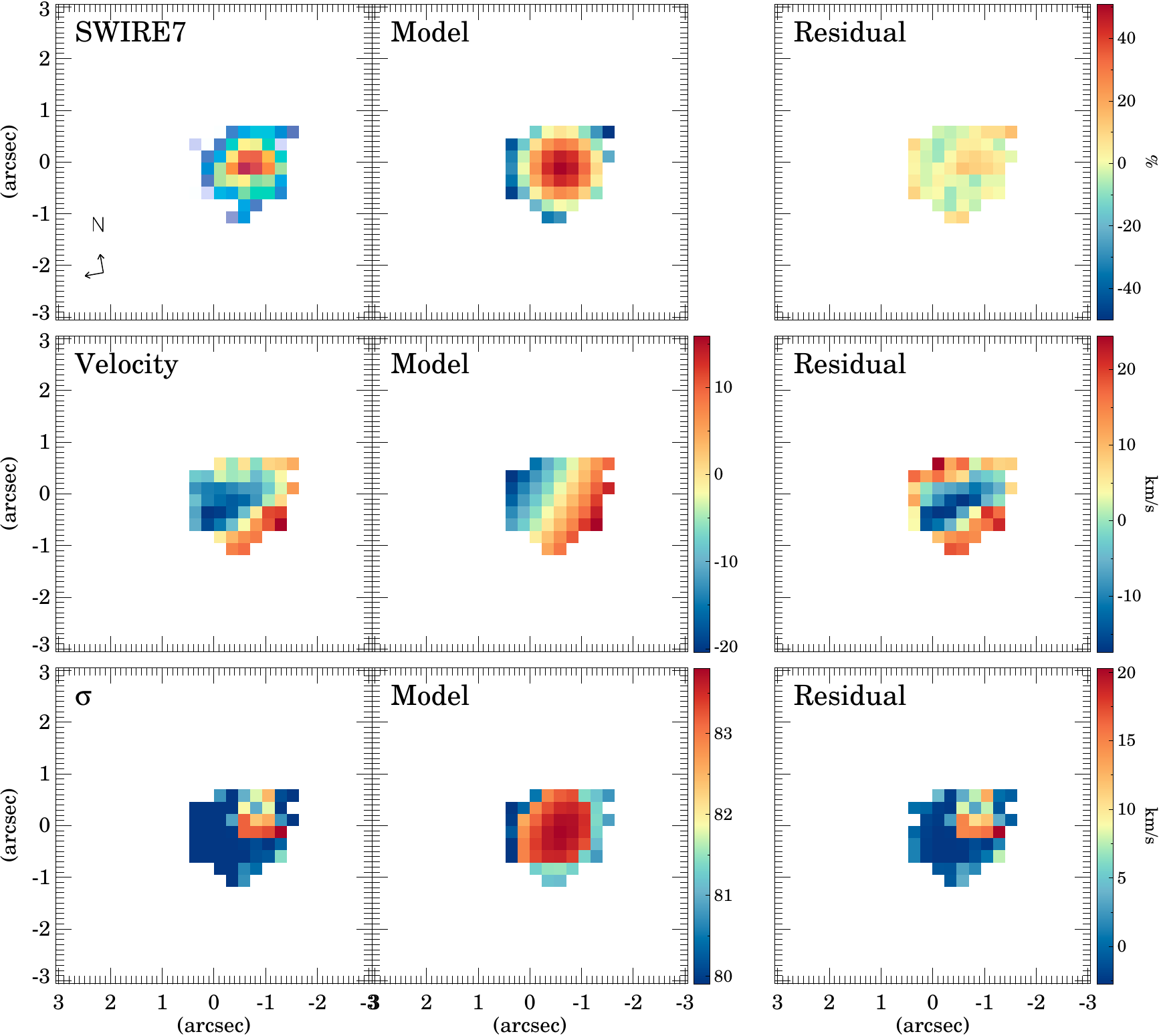}
\caption{\small Same as Figure~\ref{fig:galpak_xmm2w} but for SWIRE7.}
\end{figure*}

\begin{figure*}
\centering
\includegraphics[width=0.68\textwidth]{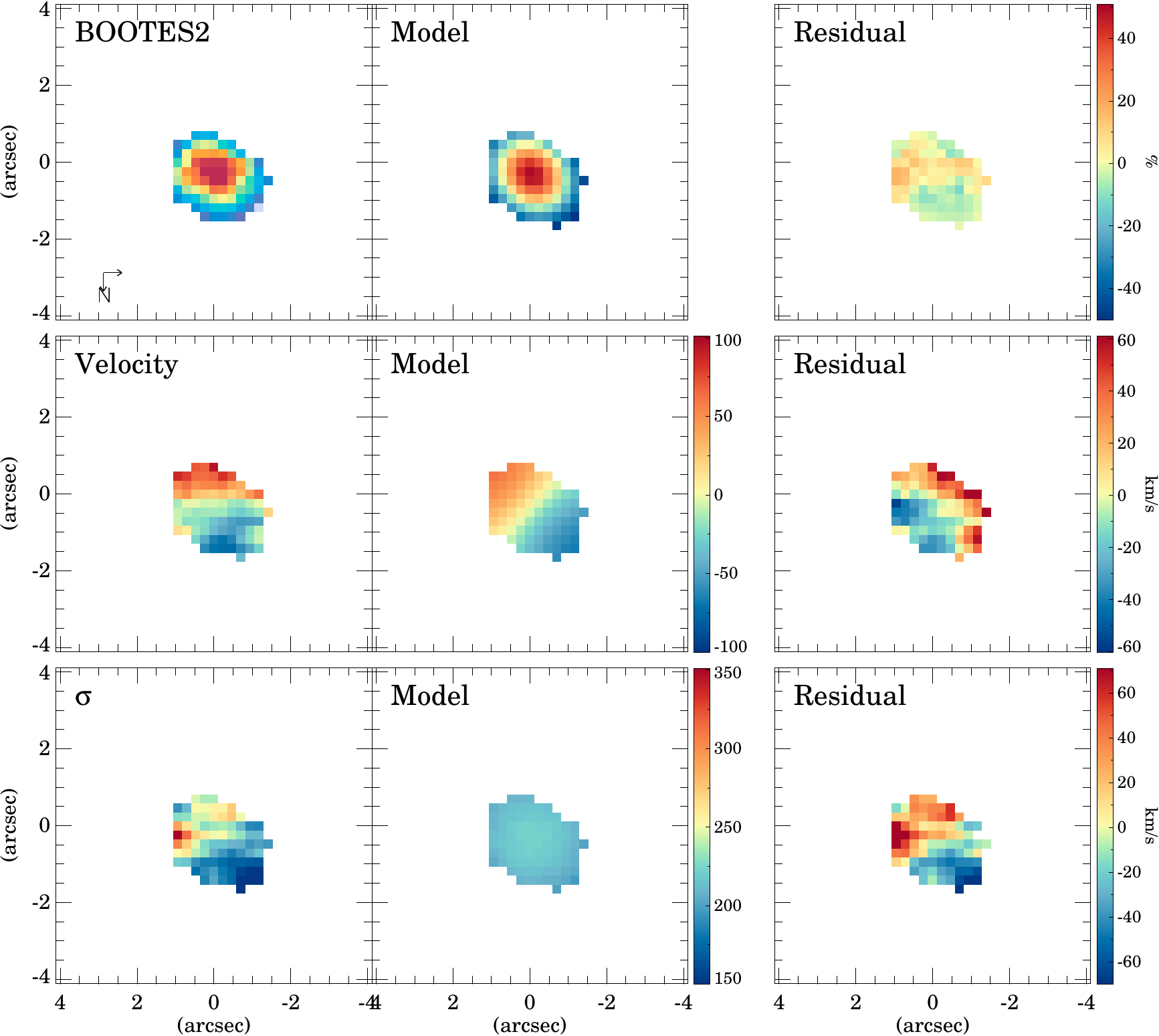}
\caption{\small Same as Figure~\ref{fig:galpak_xmm2w} but for BOOTES2.}\label{lastpage}
\end{figure*}

\end{document}